\documentclass[11pt]{article}

\usepackage[table]{xcolor}
\usepackage{times}
\usepackage{cite}
\usepackage{graphicx}
\RequirePackage{etoolbox}
\usepackage[margin=1in]{geometry}
\usepackage{amsmath,amssymb,bm,mathtools}

\usepackage[T1]{fontenc}
\usepackage[utf8]{inputenc}
\usepackage{hyperref}
\usepackage{siunitx}
\usepackage{physics}
\usepackage{microtype}
\usepackage{setspace}
\usepackage[font=small,labelfont=bf]{caption}

\newcommand{\manuscriptstretch}{1.25}

\newenvironment{sciabstract}{%
\begin{quote} \bf}
{\end{quote}}

\title{Spontaneous flows and interfacial instabilities in oxygen-sensitive living active matter}

\author
{Azam Gholami$^{1\ast}$, Sangram Gore$^{1}$, Sai V. R. Ambadipudi$^{1}$, Iraj Gholami$^{1}$,  Albert Bae$^{2}$
\\
\normalsize{$^{1}$Science Division, New York University Abu Dhabi, Abu Dhabi, UAE}\\
\normalsize{$^{2}$Lewis $\&$ Clark College, Portland, Oregon, USA.}\\
\normalsize{$^\ast$ Corresponding author: azam.gholami@nyu.edu.}
}

\date{}

\begin{document}


\setstretch{\manuscriptstretch}
\raggedbottom


\maketitle


\begin{sciabstract}
Active fluids generate motion and stress internally, but in living systems
their activity is often regulated by environmental fields that organisms
consume or produce. How such fields localise active stresses and create
flow-generating interfaces remains unclear. Here we show that oxygen
organises suspensions of the flagellated microswimmer
\textit{Euglena gracilis} into an annular living interface. In circular
chambers with an air-exposed periphery, a labyrinthine bioconvective pattern
and an annular cellular accumulation emerge nearly simultaneously, whereas
sealing the periphery suppresses the annulus. The accumulation then sharpens,
develops finite-wavelength protrusions and forms a long-lived, collectively
rotating corona. An oxygen-coupled polar active-fluid model qualitatively
recapitulates this progression: oxygen transport, cellular consumption and
oxygen-regulated swimming and reorientation assemble and position the
annulus, whereas dipolar active stresses destabilise it and generate
collective flow. These results show how a metabolically shaped chemical field
can create and activate a living interface, linking taxis, bioconvection and
active interfacial hydrodynamics.
\end{sciabstract}

%
\section*{Introduction}
Active matter comprises units that consume energy locally and convert it into
motion and mechanical stress, collectively generating flows and
spatiotemporal organisation. Suspensions of swimming microorganisms,
motor-driven cytoskeletal filaments, motile colloids and cell layers are
therefore not simply passive complex fluids, but materials maintained
intrinsically out of equilibrium
\cite{ramaswamy2010mechanics,marchetti2013hydrodynamics,
	saintillan2013active,saintillan2018rheology,ziepke2022multi,
	ziepke2025acoustic}. Their internal drive can produce spontaneous flows,
coherent vortices, propagating bands and active turbulence, raising a central
question in nonequilibrium physics: how do local motility, orientational order
and active stresses organise macroscopic fluid motion
\cite{simha2002hydrodynamic,sanchez2012spontaneous,
	wensink2012mesoscale}?

Polar active fluids are particularly rich because their constituents break
head--tail symmetry, permitting directed self-advection in addition to active
stress. This distinguishes wet polar suspensions from active nematics and
connects them to the hydrodynamics of polar order and flocking
\cite{toner1995long,toner2005hydrodynamics}. In suspensions and active gels,
active stresses can destabilise ordered states and generate flow, while polar
self-advection controls the propagation and rotation of the resulting
patterns. These instabilities depend sensitively on orientational order,
confinement, boundary conditions and the rheological response of the active
material
\cite{ramaswamy2007active,kruse2004asters,voituriez2005spontaneous,
	voituriez2006generic,saintillan2007orientational,
	saintillan2008instabilities,giomi2008complex,giomi2012polar,
	duan2023dynamical,gompper20202020}. Experiments have demonstrated polar
streams, flocks, confined vortices and macroscopic rotating states
\cite{schaller2010polar,schaller2011polar,bricard2013emergence,
	bricard2015emergent,geyer2018sounds,zhang2022polar}. Most theoretical
descriptions, however, treat motility and activity as prescribed material
parameters. In living suspensions, both can instead be regulated by
environmental fields that the organisms consume, produce or transport.

Oxygen is one such dynamical field. It is supplied at air--liquid interfaces,
transported by diffusion and flow, and consumed by cells. Oxygen gradients
have long been known to organise swimming microorganisms through oxytaxis and
bioconvection
\cite{platt1961bioconvection,pedley1992hydrodynamic,
	hill2005bioconvection,hillesdon1996bioconvection,dombrowski2004self,
	tuval2005bacterial,hokmabad2025spatial}. More recent work has shown that
self-generated oxygen gradients can control aggregation in dense
photosynthetic suspensions, and that oxygen depletion or air-impermeable
confinement can induce marked transitions in algal bioconvection
\cite{fragkopoulos2021self,gore2025oxygen}. These studies establish oxygen as
a control field for living suspensions, but leave open what happens when an
oxygen-selected cellular accumulation enters a strongly collective regime
capable of supporting active stress and macroscopic flow.

The flagellated protist \textit{Euglena gracilis} provides a living polar
fluid in which to address this question. Individual cells swim using a single
anterior flagellum and exhibit complex trajectories, adaptive phototaxis and
flagellar beat switching
\cite{rossi2017kinematics,tsang2018polygonal}. They also respond strongly to
oxygen. Early experiments showed that apparent red-light-induced accumulation
in green \textit{Euglena} is largely mediated by photosynthetic oxygen
production rather than by direct red-light phototaxis
\cite{checcucci1974red}. Subsequent studies implicated cytochrome
\textit{c} oxidase in oxygen sensing and demonstrated that
\textit{Euglena} suspensions can form dynamic ring patterns through
chemosensory responses to dissolved oxygen
\cite{miller1978cytochrome,colombetti1978chemosensory,
	porterfield1997oxytaxis}. Oxygen-induced annular migration is not unique to
\textit{Euglena}: oxygen consumption by confined
\textit{Dictyostelium discoideum} colonies can likewise drive collective
aerotactic migration into dense rings
\cite{cochet2021hypoxia}. Earlier observations of phase-boundary-induced
patterns further illustrate the sensitivity of \textit{Euglena} to
interfaces and confinement
\cite{brinkmann1968phasengrenzen}. Hydrodynamically,
\textit{E. gracilis} behaves, on average over a flagellar beat, as an
off-axis puller \cite{giuliani2021how}. Dense suspensions of this organism
therefore combine polar swimming, oxygen-regulated motility, active stresses
and environmental feedback.

Here we show that oxygen-dependent taxis can assemble a living active
interface that subsequently undergoes a flow-generating instability. In
circular chambers with an air-exposed periphery, peripheral oxygen supply and
cellular consumption establish a predominantly radial oxygen organisation
during the early evolution. Starting from an initially homogeneous
suspension, an annular cellular accumulation emerges at the outer edge of a
developing labyrinthine bioconvective pattern, with no clearly resolved delay
between their onsets. The annular accumulation subsequently sharpens and
develops finite-wavelength, corona-like protrusions accompanied by sustained
collective circulation. Sealing the peripheral boundary suppresses formation
of a comparable annulus, whereas a ring can form near an oxygen source in a
strongly confined chamber without developing a rotating corona. These
observations separate oxygen-guided localisation from the later mechanical
instability: oxygen-dependent taxis assembles and radially positions the
cellular interface, whereas collective activity destabilises it.

The corona connects these observations to the broader physics of active
interfaces, active droplets and confined active gels. Active stresses can
destabilise material and phase boundaries, generate interfacial waves and
giant fluctuations, drive spontaneous droplet motion and produce protrusive
or shape-changing morphologies
\cite{tjhung2012spontaneous,soni2019stability,adkins2022dynamics,
	tayar2023controlling,xu2023geometrical,gulati2024traveling}. In active
droplets and confined polar materials, distortions of the director or
polarisation can couple to interfacial deformation and flow: contractile
systems can undergo splay-mediated symmetry breaking, while anchoring and
confinement can select rotating states and complex morphologies
\cite{tjhung2012spontaneous,fialho2017anchoring,keber2014topology,
	nejad2023spontaneous,marenduzzo2007steady,edwards2009spontaneous,bhattacharyya2025active}. In these systems, however, the interface is usually
provided by phase separation, material contrast or an externally imposed
boundary. Here the interface is assembled by the cells themselves through
their response to an oxygen field that they simultaneously modify. The
observed corona is therefore an instability of a chemically selected living
interface.

To test this interpretation, we develop an oxygen-coupled polar active-fluid
model based on the Giomi--Marchetti framework
\cite{giomi2012polar}. The model couples cell concentration, polarisation,
fluid velocity and dissolved oxygen. Oxygen enters through two biologically
motivated effects: a sign-changing oxytactic torque, which reorients the
polarisation towards or away from oxygen gradients, and an oxygen-dependent
swimming speed, which suppresses motility in oxygen-poor regions. Simulations
initialised with a homogeneous concentration field and zero fluid velocity
qualitatively recapitulate the progression from annular localisation to a
protrusive, flowing state. Oxygen-regulated motility and oxytaxis generate the
annular band, whereas sufficiently large dipolar activity destabilises the
resulting interface and produces finite-wavelength protrusions and azimuthal
flow. The simulated polarisation develops strong spatial heterogeneity,
including splay and bend distortions and a loss of coherent tangential order,
showing that the instability is not a purely density-driven deformation.

The model also separates the onset of circulation from the selection of its
direction. The simulations impose a signed tangential Dirichlet condition for
the polarisation at the peripheral boundary. Reversing the initial bulk
polarisation while retaining the boundary polarity changes the transient
dynamics but not the long-time circulation, whereas reversing the boundary
polarity reverses the final flow. Thus, active stresses generate the
instability and circulation, but the signed tangential anchoring selects its
chirality; the predominantly radial oxygen field does not itself impose a
clockwise or counter-clockwise bias.

Our results identify a mechanism by which an environmental field can
organise, position and activate a living fluid. Oxygen transport, consumption
and taxis assemble a cellular interface; active stresses acting through the
reorganised polarisation field destabilise that interface and generate
macroscopic flow; and polar boundary anchoring selects the direction of the
resulting circulation. This coupling between metabolism, taxis,
orientational order and active hydrodynamics provides a route by which living
systems can generate chemically organised active interfaces and suggests
strategies for controlling active-matter flows through chemical and geometric
design.
\section*{Results}
\subsection*{Active stresses destabilise an oxytactically formed cellular ring}
\begin{figure}[!htbp]
	\centering
	\includegraphics[width=\columnwidth]{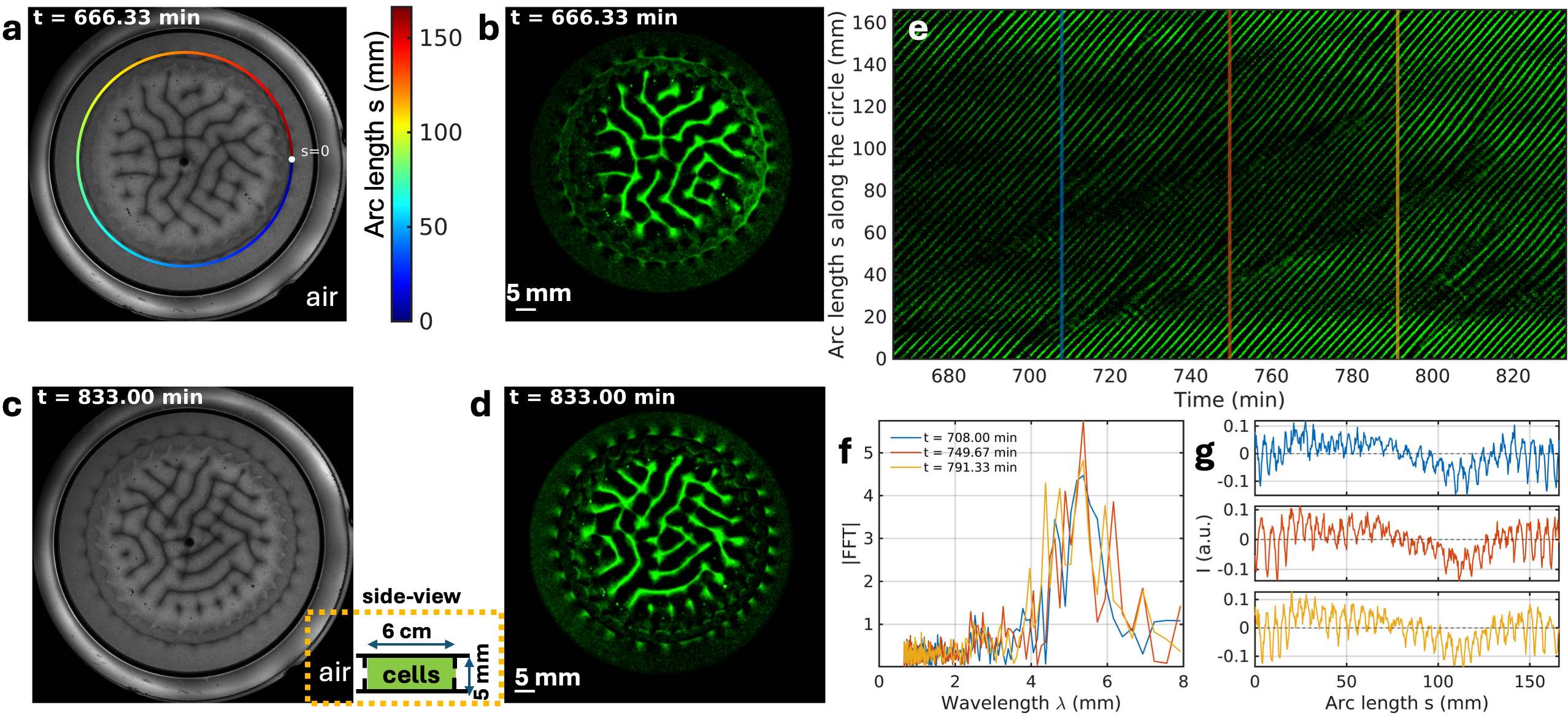}
	\caption{\textbf{Spontaneous formation and rotation of a cellular corona.}
		\textbf{a--d,} Time-lapse images of an initially homogeneous
		suspension of \textit{E. gracilis} in a circular chamber with an
		air-exposed periphery. Raw images (\textbf{a,c}) show the emergence
		of dynamic bioconvection and an annular cellular accumulation; the
		corresponding processed false-colour images are shown in
		\textbf{b,d}. The side view in \textbf{c} shows the unsupported
		peripheral meniscus forming the air--liquid boundary. The annular
		accumulation subsequently develops rotating corona-like protrusions.
		Scale bars in \textbf{b,d}, \(5~\mathrm{mm}\); chamber diameter in
		the side view, \(6~\mathrm{cm}\).
		\textbf{e,} Space--time plot constructed by stacking intensity
		profiles sampled along the circular path indicated in \textbf{a},
		showing persistent angular propagation of the protrusive pattern.
		\textbf{f,} Fourier spectra of the intensity profiles in
		\textbf{g}, yielding a characteristic protrusion wavelength of
		approximately \(5~\mathrm{mm}\).
		\textbf{g,} Representative intensity profiles sampled at the blue,
		red and orange time points indicated in \textbf{e}.}
	\label{fig:corona}
\end{figure}

We introduced \textit{E. gracilis} suspensions at an initial density of
\((5\text{--}8)\times10^6~\mathrm{cells\,mL^{-1}}\) into circular
polymethyl methacrylate (PMMA) chambers with a diameter of
\(6~\mathrm{cm}\), a height of \(5~\mathrm{mm}\), and an air-exposed
periphery. The initially homogeneous cell distribution rapidly reorganised:
a labyrinthine bioconvective pattern developed in the chamber interior while,
almost simultaneously, a dense annular accumulation emerged at its outer edge
(Fig.~\ref{fig:corona}a--d and Supplementary Video~1). The ring and
labyrinth subsequently sharpened and evolved together, with no clearly
resolved delay between their onsets. A comparable annular accumulation was
observed only in chambers exposed to air at the periphery. In this circular
geometry, oxygen supplied at the outer boundary and consumed throughout the
suspension is expected to establish a predominantly radial oxygen gradient,
with oxygen concentration generally highest near the air-exposed perimeter
and decreasing towards the chamber centre. These observations are therefore
consistent with oxygen taxis selecting the radial position of the dense
annular accumulation during the concurrent development of the labyrinthine
pattern.

As a control, we performed experiments in PMMA chambers that were completely
sealed at the periphery, thereby suppressing oxygen exchange with the
surrounding air. Under these conditions, no comparable annular accumulation
was observed, supporting the conclusion that peripheral oxygen access is
required for ring formation (Fig.~\ref{fig:control} and Supplementary
Video~2). This oxygen-mediated interpretation is consistent with earlier
observations that \textit{E. gracilis} can form dynamic ring patterns in thin
chambers through chemosensory responses to dissolved oxygen rather than
passive cell aggregation~\cite{colombetti1978chemosensory}. Direct experiments
with an externally imposed oxygen gradient further demonstrated positive
oxytaxis: dark-adapted \textit{E. gracilis} swam towards higher
\(\mathrm{O}_2\) concentrations and accumulated in a dense band at the oxygen
front~\cite{porterfield1997oxytaxis}. Earlier work also showed that apparent
red-light-induced accumulation in green \textit{Euglena} is largely mediated
by photosynthetic oxygen production, further supporting the role of oxygen as
a motility cue in this organism~\cite{checcucci1974red}.

As the annular accumulation intensified and became more sharply defined, the
ring behaved as a dynamically active interface. It subsequently developed
corona-like protrusions and underwent persistent collective rotation
(Fig.~\ref{fig:corona}a--e). This transition indicates that the corona is not
determined solely by the oxygen gradient, but instead arises from a dynamical
instability of the oxytactically assembled ring. A space--time plot constructed
from intensity profiles sampled along the annulus revealed sustained
propagation of the protrusive pattern (Fig.~\ref{fig:corona}e). Fourier
analysis of these intensity profiles yielded a characteristic protrusion
wavelength of approximately \(5~\mathrm{mm}\)
(Fig.~\ref{fig:corona}f,g). Both the rotational motion and the protrusions
persisted over long times, underscoring the sustained active nature of the
instability.

\begin{figure}[!htbp]
	\centering
	\includegraphics[width=\columnwidth]{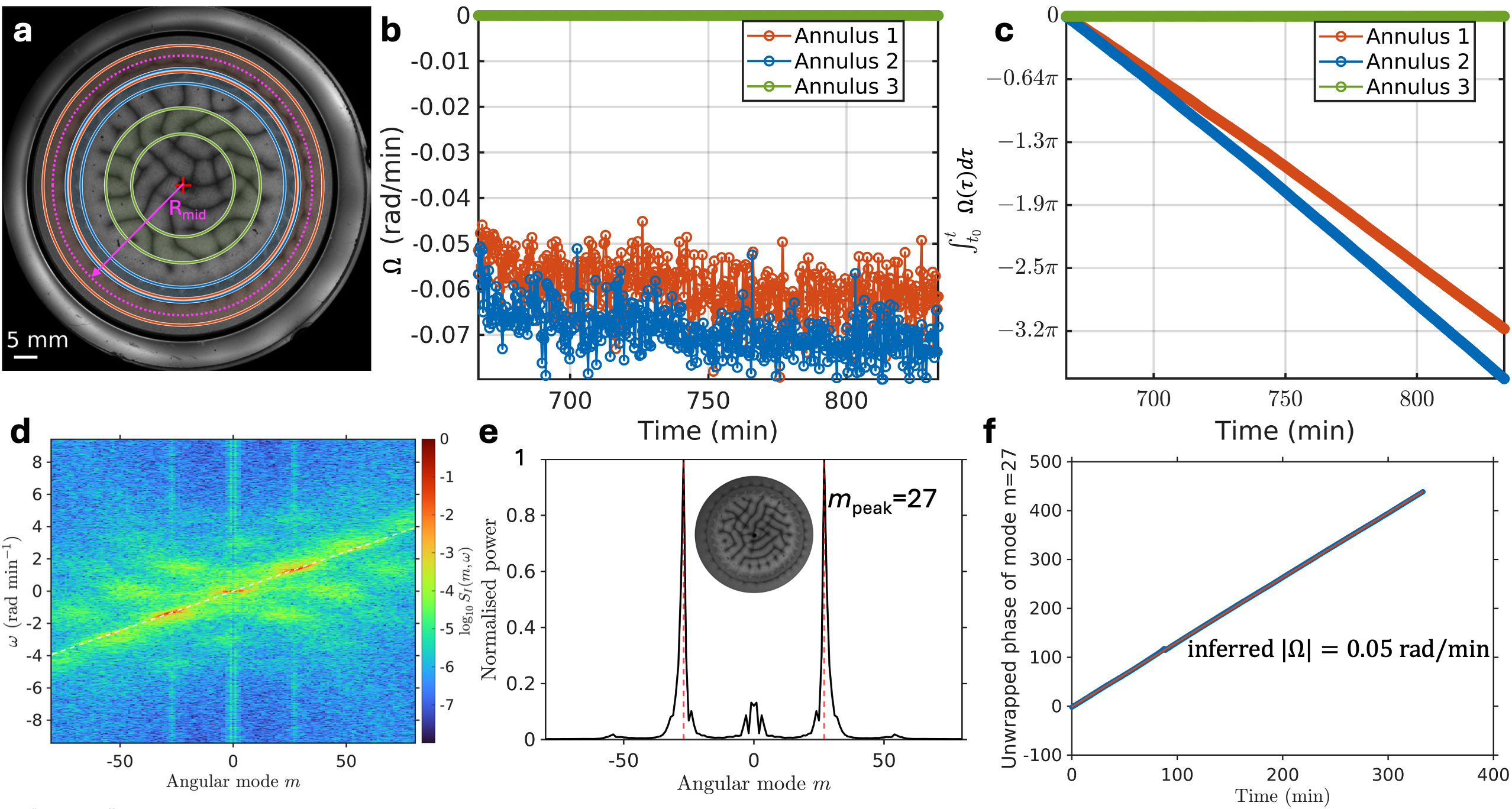}
	\caption{\textbf{Angular dynamics and spectral signature of corona
			rotation.}
		\textbf{a}, Representative image showing the annular regions used for
		polar-coordinate cross-correlation. The orange and blue annuli encompass
		the corona, whereas the green annulus is an inner control region. The red
		cross marks the centre of the polar transformation, and the magenta dashed
		circle marks the mean analysis radius \(R_{\mathrm{mid}}\). 
		\textbf{b}, Instantaneous angular rotation rate \(\Omega(t)\) obtained by
		cross-correlating successive intensity patterns. With \(\theta\) increasing
		counter-clockwise, \(\Omega<0\) denotes clockwise rotation. The two outer
		annuli rotate persistently clockwise, whereas the inner annulus shows little
		net motion.
		\textbf{c}, Cumulative angular displacement,
		\(\Theta(t)=\int_{t_0}^{t}\Omega(\tau)\,\mathrm{d}\tau\), for the three
		annuli.
		\textbf{d}, Laboratory-frame angular dynamic structure factor
		\(S_I(m,\omega)\) of the radially averaged annular intensity fluctuations,
		shown on a logarithmic scale. The oblique spectral ridge indicates angular
		propagation, and the dashed line shows the drift relation
		\(\omega=-m\Omega\).
		\textbf{e}, Normalised angular-mode power spectrum obtained by integrating
		\(S_I(m,\omega)\) over frequency. Red dashed lines mark the symmetric
		dominant peaks at \(m=\pm27\). The positive branch,
		\(m_{\mathrm{peak}}=27\), corresponds to an azimuthal wavelength
		\(\lambda=2\pi R_{\mathrm{mid}}/\lvert m_{\mathrm{peak}}\rvert
		\simeq5~\mathrm{mm}\). The inset shows the analysed corona and its mean
		radius.
		\textbf{f}, Unwrapped phase \(\phi_{27}(t)\) of the positive dominant mode
		(blue) and linear fit (orange). Using
		\(\Omega=-(1/m)\,\mathrm{d}\phi_m/\mathrm{d}t\), the fitted phase slope gives
		\(\Omega\simeq-0.05~\mathrm{rad\,min^{-1}}\), consistent with clockwise
		rotation.}
	\label{fig:corona2}
\end{figure}

We first quantified the collective rotation in real space using
polar-coordinate cross-correlation. Intensity patterns at successive time
points were compared within the annular regions shown in
Fig.~\ref{fig:corona2}a, yielding the frame-to-frame angular displacement
and the corresponding instantaneous angular velocity \(\Omega(t)\)
(Fig.~\ref{fig:corona2}b). With the convention that \(\theta\) increases
counter-clockwise, negative values of \(\Omega\) correspond to clockwise
rotation. The two outer annuli, which encompass the corona, displayed
persistent clockwise motion, whereas the inner control annulus showed little
net rotation. Integrating \(\Omega(t)\) yielded the cumulative angular
displacement, which decreased approximately linearly with time and thereby
confirmed the sustained clockwise rotation of the protrusive corona
(Fig.~\ref{fig:corona2}c).

As an independent spectral characterisation of the same dynamics, we computed
an angular dynamic structure factor from the annular intensity field. The
intensity was radially averaged over the corona region to obtain
\(I_{\mathrm{ann}}(\theta,t)\), and the angular fluctuation
\[
\delta I_{\mathrm{ann}}(\theta,t)
=
I_{\mathrm{ann}}(\theta,t)
-
\langle I_{\mathrm{ann}}(\theta,t)\rangle_\theta
\]
was Fourier transformed in angle and time. The resulting laboratory-frame
spectrum \(S_I(m,\omega)\) displayed an oblique ridge, indicating angular
propagation of the corona pattern (Fig.~\ref{fig:corona2}d). The ridge was
consistent with the drift relation expected for collective rotation,
\(\omega \simeq -m\Omega\). Integrating \(S_I(m,\omega)\) over frequency
yielded an angular-mode power spectrum with a dominant finite mode at
\(m_{\mathrm{peak}}=27\) (Fig.~\ref{fig:corona2}e). This mode corresponds to
an azimuthal protrusion wavelength
\[
\lambda
=
\frac{2\pi R_{\mathrm{mid}}}{m_{\mathrm{peak}}},
\]
where \(R_{\mathrm{mid}}\) is the mean radius of the analysed annulus. The
result is consistent with the approximately \(5~\mathrm{mm}\) wavelength
measured directly from the real-space intensity profiles in
Fig.~\ref{fig:corona}f,g. Phase tracking of the dominant mode gave an
approximately linear phase evolution and a mean rotation speed
\(|\Omega|\simeq0.05~\mathrm{rad\,min^{-1}}\)
(Fig.~\ref{fig:corona2}f), in agreement with the polar-coordinate
cross-correlation analysis.

Together, the real-space cross-correlation and angular dynamic structure
factor establish that the corona is a rotating, finite-wavelength
interfacial pattern. Combined with the oxygen-control experiments and
numerical simulations, these measurements support a two-stage
interpretation: oxygen-dependent taxis assembles and radially positions the
annular cellular accumulation, whereas active stresses destabilise the
resulting interface, generating protrusive deformation and sustained
collective rotation.
\subsubsection*{Long-time evolution and confined controls separate oxytaxis from corona formation}
\begin{figure}[!htbp]
	\centering
	\includegraphics[width=\columnwidth]{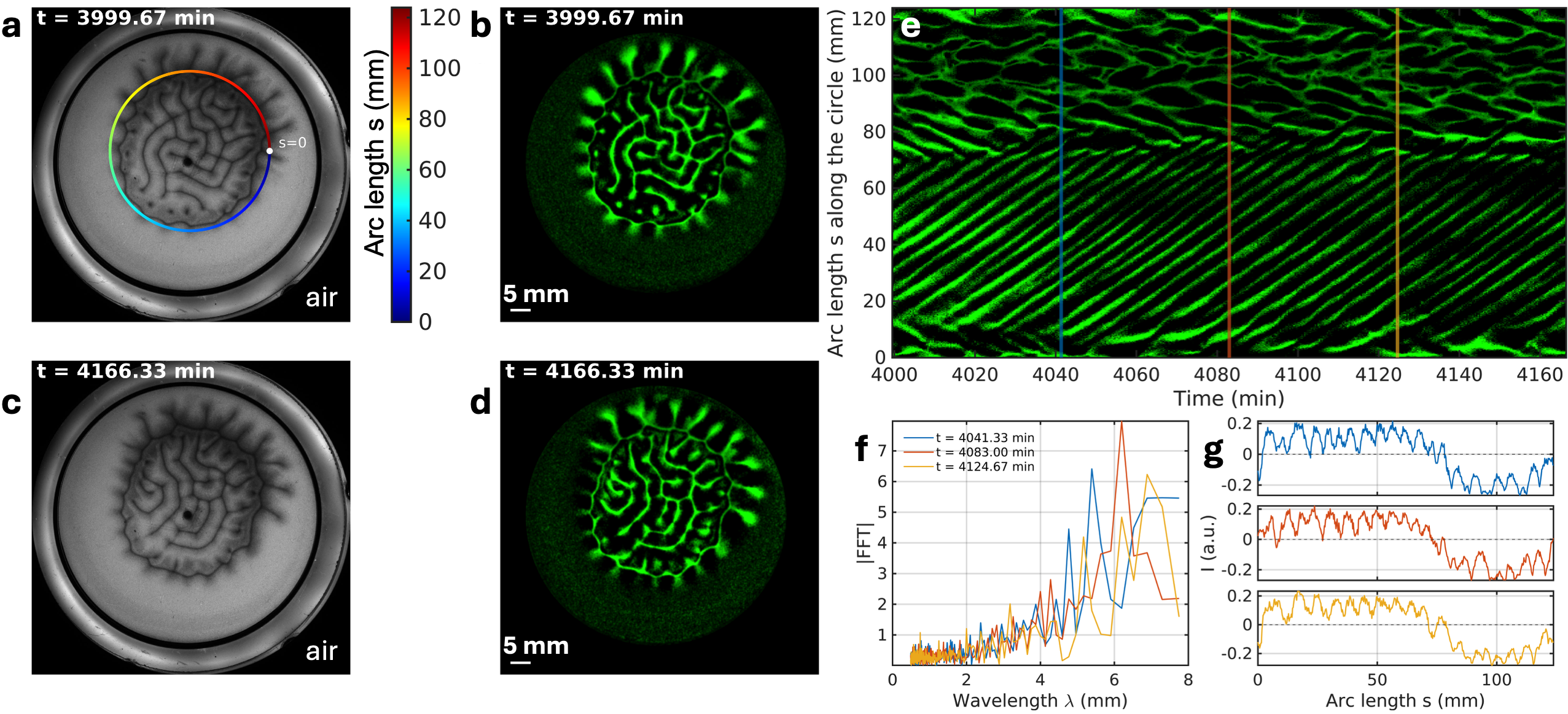}
	\includegraphics[width=0.96\columnwidth]{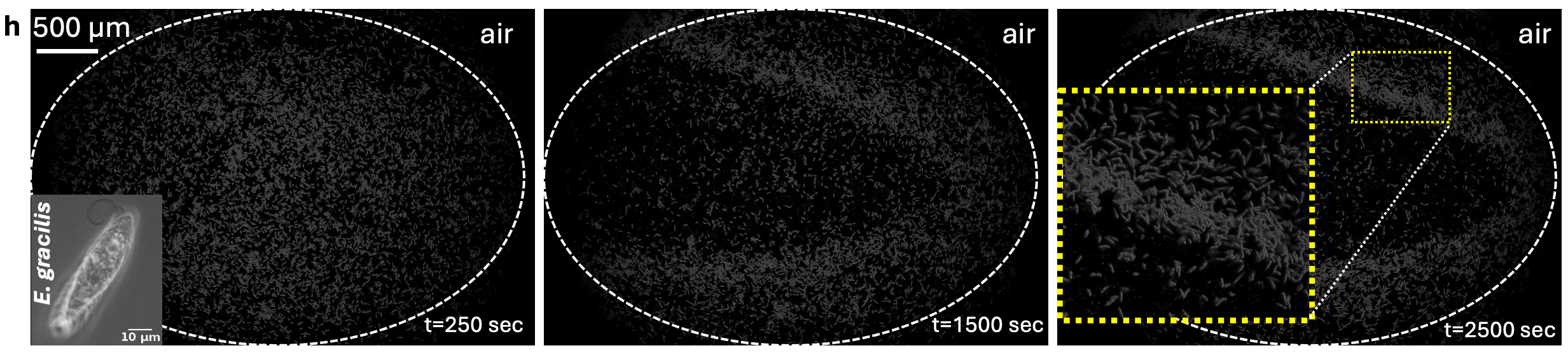}
	\caption{\textbf{Long-time evolution from a rotating corona to large
			dynamic protrusions.}
		\textbf{a--d,} Time-lapse images from an experiment lasting more than
		\(2~\mathrm{d}\), showing the transition from a coherently rotating
		corona to larger, irregular protrusions. Raw images are shown in
		\textbf{a,c}, with the corresponding processed false-colour images in
		\textbf{b,d}. The annular band contracts slightly as global rotational
		order is progressively lost. 
		\textbf{e,} Space--time plot constructed along the circular path
		indicated in \textbf{a}. Angular propagation persists over the sector
		\(0\lesssim s\lesssim60~\mathrm{mm}\), whereas protrusions elsewhere
		undergo lateral oscillations and branching.
		\textbf{f,} Fourier spectra of the representative intensity profiles
		in \textbf{g}, showing a characteristic wavelength of approximately
		\(5~\mathrm{mm}\).
		\textbf{g,} Intensity profiles sampled at the blue, red and orange
		time points indicated in \textbf{e}.
		\textbf{h,} Time-lapse images from a confined control experiment in
		which the suspension was held between two microscope slides separated
		by \(h=100~\mu\mathrm{m}\), with an air pocket on the right. A cellular
		ring forms near the air--liquid interface, but bioconvective motion and
		the dynamic protrusive instability are not observed. }
	\label{fig:longtime}
\end{figure}

At longer times, the rotating corona evolved into a second dynamical regime
(Fig.~\ref{fig:longtime}). In experiments lasting more than \(2~\mathrm{d}\), the initially
coherent rotational motion progressively lost global order, the circular ring
contracted slightly, and the corona-like pattern evolved into larger,
highly dynamic protrusions (Fig.~\ref{fig:longtime}a--e). A space--time plot
measured along the circular path indicated in Fig.~\ref{fig:longtime}a
revealed heterogeneous dynamics around the annulus. Rotational motion persisted
over one sector of the ring, corresponding to arc lengths
\(0 \lesssim s \lesssim 60~\mathrm{mm}\), whereas protrusions in the remaining
sector displayed beat-like lateral oscillations and branching events
(Fig.~\ref{fig:longtime}e). Fourier analysis of intensity profiles extracted
from the kymograph again gave a dominant wavelength of approximately
\(5~\mathrm{mm}\) (Fig.~\ref{fig:longtime}f,g), indicating that the characteristic length
scale of the protrusive instability is retained even as the long-time dynamics
becomes more heterogeneous.

We next tested whether oxytactic localisation alone is sufficient to generate
the dynamic corona. In a separate confined experiment, cells were held between
two microscope slides with a gap height of \(h=100~\mu\mathrm{m}\), with an
air pocket on one side of the chamber. Under these conditions, a cellular ring
formed near the oxygen source, consistent with previous reports of
oxygen-guided accumulation and ring formation in \textit{Euglena}
\cite{colombetti1978chemosensory,brinkmann1968phasengrenzen}. However, the
reduced height suppressed bioconvection, and no dynamic interfacial
instability was observed (Fig.~\ref{fig:longtime}h). This confined geometry
therefore separates the two processes: oxygen taxis can localise the cells into
a ring, whereas the rotating corona requires a sufficiently developed active
interface capable of supporting collective flow and mechanical
destabilisation.
\subsubsection*{Oxygen-coupled polar active-fluid simulations}
\begin{figure}[!htbp]
	\centering
	\includegraphics[width=\columnwidth]{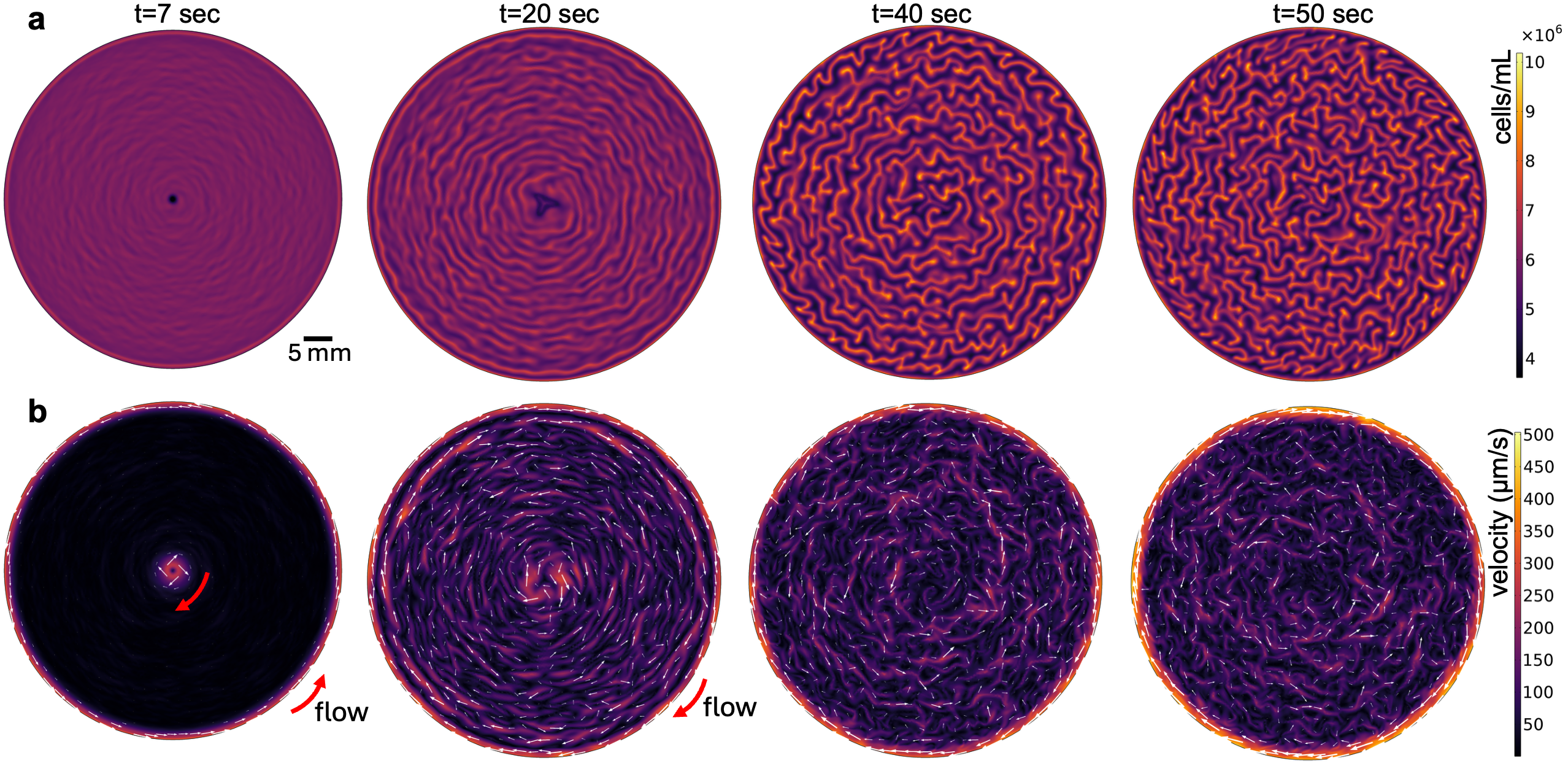}
	\includegraphics[width=\columnwidth]{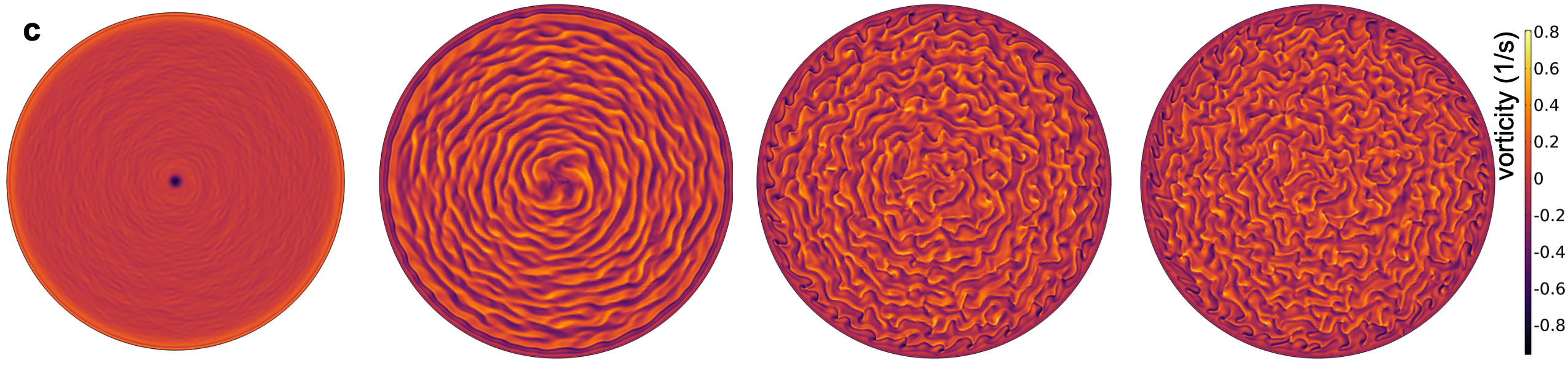}
	\caption{\textbf{Simulated evolution of the cellular concentration,
			flow and vorticity.}
		\textbf{a,} Cell-concentration fields at
		\(t=7,\,20,\,40\) and \(50~\mathrm{s}\), from left to right.
		An annular accumulation and concentric concentration modulations
		develop and progressively reorganise into a chamber-spanning
		labyrinthine pattern. 
		\textbf{b,} Corresponding maps of the fluid-speed magnitude,
		\(|\mathbf{u}|\), with white arrows showing the local velocity
		direction. Red arrows indicate the sense of the large-scale
		circulation. At the earliest time, the inner and peripheral regions
		circulate in opposite directions, forming a counter-rotating flow
		configuration. As the concentration pattern evolves, the flow
		reorganises and the clockwise azimuthal component becomes dominant
		over most of the chamber, although local vortical fluctuations remain.
		\textbf{c,} Corresponding signed out-of-plane vorticity,
		\(\omega_z=\partial_xu_y-\partial_yu_x\), showing the evolution from
		the initial counter-rotating structure to a spatially heterogeneous
		vortical state. Colour scales are held fixed across all time points.
		For this simulation,
		\(\alpha=2\times10^8~\mathrm{N\,m\,mol^{-1}}\); all other parameters
		are listed in Table~S1.}
	\label{fig:model}
\end{figure}

To test this mechanism, we numerically integrated the oxygen-coupled polar
active-fluid equations (see Methods). The simulations were initialised with a
homogeneous cell concentration, zero fluid velocity and a weakly perturbed
tangential polarisation field. Along the air-exposed outer boundary, we fixed
the dissolved-oxygen concentration at its saturation value and imposed a
signed tangential Dirichlet condition with clockwise polarity,
$
\left.\mathbf P\right|_{\partial\Omega}
=
P_{\mathrm{wall}}\hat{\mathbf t}_{\mathrm{CW}}.
$
This condition fixes both the tangential orientation and the polarity of
\(\mathbf P\) at the boundary. During the early localisation stage,
peripheral oxygen supply, transport through the chamber and cellular
consumption established a predominantly radial oxygen gradient.
Oxygen-regulated swimming and the sign-changing oxytactic response then
generated convergent radial cell transport, producing an annular concentration
band (Fig.~\ref{fig:model}a). The ring was therefore not prescribed in the
initial concentration field, but emerged dynamically from the coupled effects
of oxygen transport and consumption, oxygen-regulated motility and oxytaxis.

At later times, the annular concentration band sharpened and entered a
dynamically active regime. For sufficiently large dipolar activity,
\(\alpha>\alpha_c\), active stresses amplified azimuthal perturbations of the
initially near-circular band, generating corona-like protrusions with a
characteristic finite wavelength. The protrusions propagated around the
annulus and were accompanied by a large-scale azimuthal flow. Initially, the
inner and peripheral regions circulated in opposite directions, producing a
counter-rotating flow configuration. As the instability developed, clockwise
circulation became dominant across most of the chamber, although local
vortical fluctuations persisted (Fig.~\ref{fig:model}b,c). Below the activity
threshold, the oxytactically assembled annulus remained approximately
axisymmetric and exhibited neither pronounced protrusions nor sustained
coherent circulation. The peak simulated fluid speed reached approximately
\(500~\mu\mathrm{m\,s^{-1}}\), about five times the characteristic swimming
speed of an individual \textit{E. gracilis} cell,
\(v_{\mathrm{cell}}\simeq100~\mu\mathrm{m\,s^{-1}}\)
(Fig.~\ref{fig:model}b). These results identify protrusive deformation and
collective circulation as coupled manifestations of the same active
interfacial instability.

The direction of the long-time circulation was selected by the signed
tangential boundary condition. Reversing the initial bulk polarisation while
retaining the clockwise boundary polarity modified the early transient but
did not reverse the long-time circulation, which remained clockwise. In a
complementary simulation in which the boundary polarity and a compatible
initial polarisation were both reversed, the final circulation also reversed.
These controls indicate that, within the parameter regime studied here, the
signed tangential anchoring is the dominant selector of the long-time flow
direction, whereas the initial bulk polarisation primarily affects the
transient dynamics. The predominantly radial oxygen field therefore does not
itself impose a chiral bias. This absence of an oxygen-selected circulation
direction is also consistent with experiments in which different sectors of
the same corona rotate in opposite directions
(Fig.~\ref{fig:corona_BothDirections}).
\begin{figure}[!htbp]
	\centering
	\includegraphics[width=\columnwidth]{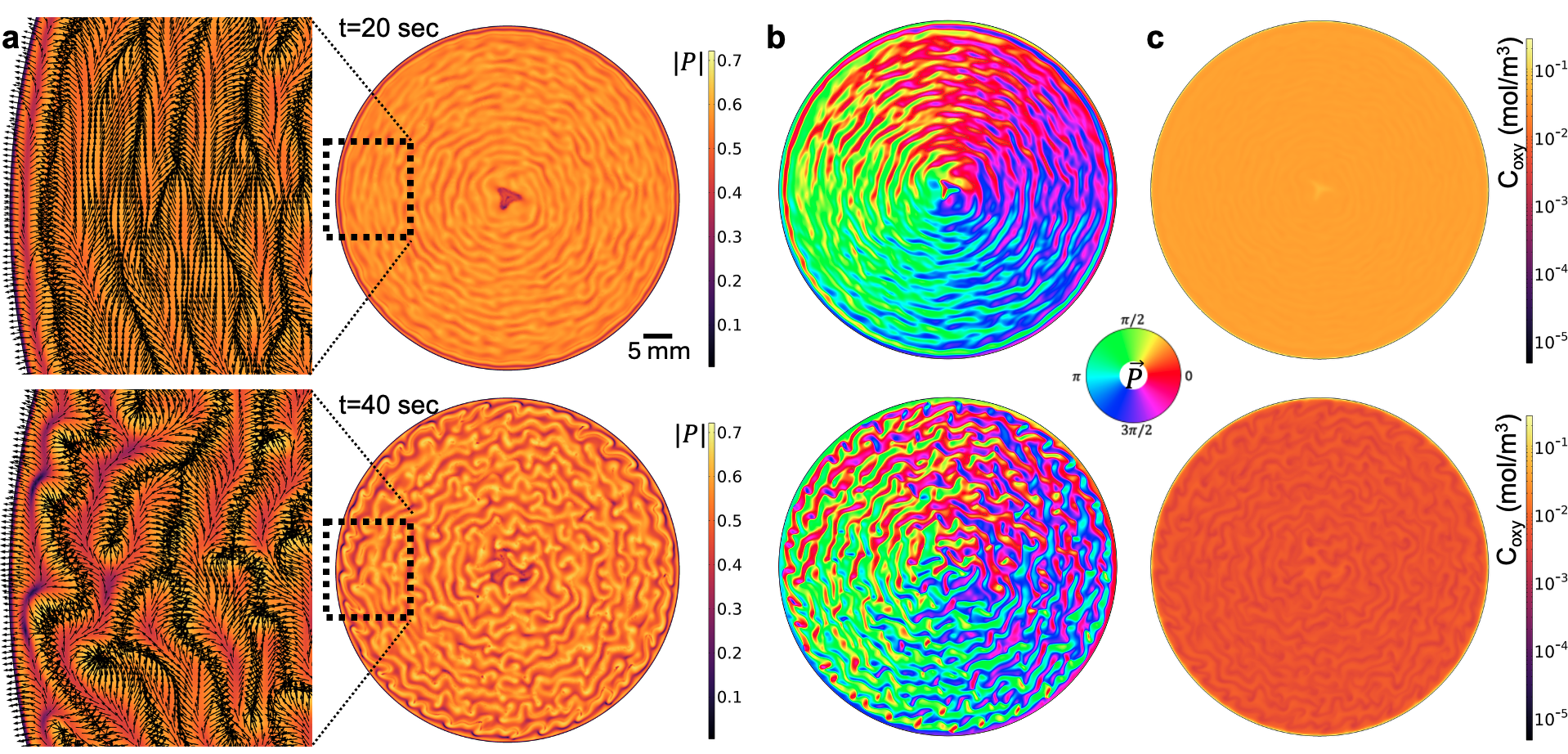}
	\includegraphics[width=\columnwidth]{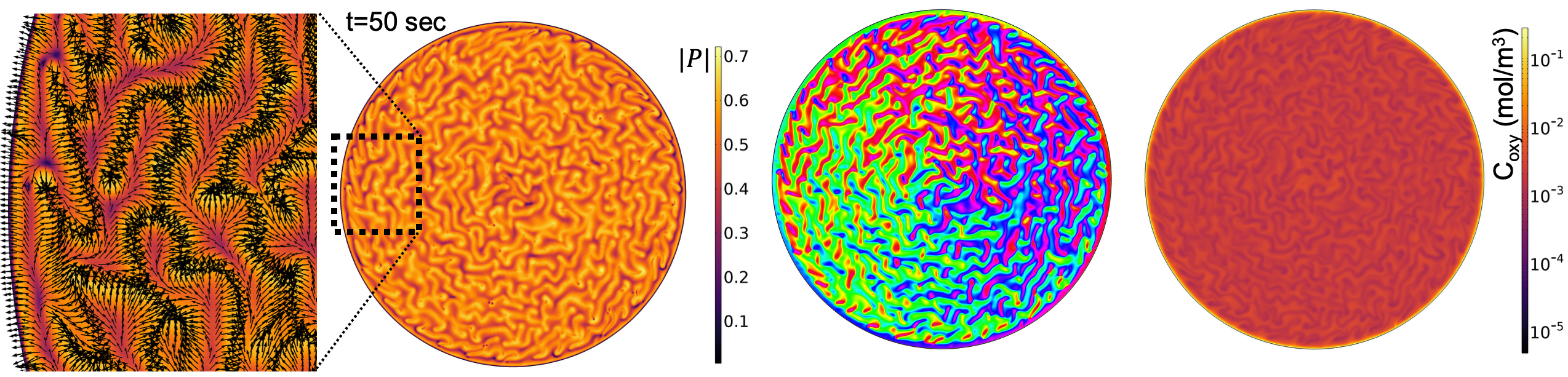}
	\caption{\textbf{Polarisation reorganisation and oxygen-field evolution
			during the activity-driven instability.}
		\textbf{a,} Simulated polarisation magnitude,
		\(|\mathbf P|\), at \(t=20\), \(40\) and \(50~\mathrm{s}\), from
		top to bottom. Bright bands indicate enhanced polar order, whereas
		dark regions indicate reduced polarisation magnitude. Although the
		colour scale represents \(|\mathbf P|\), the bright bands spatially
		coincide with the high-cell-concentration ridges shown at the
		corresponding times in Fig.~\ref{fig:model}a. Dashed boxes mark the
		regions displayed in the enlarged views, where black arrows show the
		local polarisation direction and reveal the progressive reorganisation
		of the initially coherent polar texture into strongly distorted
		domains associated with the developing cellular pattern. 
		\textbf{b,} Corresponding maps of the polarisation phase,
		\(\theta_P=\operatorname{mod}
		[\operatorname{atan2}(P_y,P_x),2\pi]\), shown using a cyclic colour
		map over \(0\leq\theta_P<2\pi\). The colour wheel represents the
		local orientation of the polar vector. Because \(\mathbf P\) is a
		true polar field, antiparallel orientations \(\mathbf P\) and
		\(-\mathbf P\) are distinct and are represented by colours separated
		by \(\pi\). The phase maps show the progressive breakup of the
		initially coherent orientation field into spatially heterogeneous
		polar domains.
		\textbf{c,} Corresponding dissolved-oxygen concentration,
		\(C_{\mathrm{oxy}}\), shown on a logarithmic colour scale. The oxygen
		field becomes increasingly modulated as the cellular and
		polarisation patterns develop.}
	\label{fig:polar_order}
\end{figure}
\begin{figure}[!htbp]
	\centering
	\includegraphics[width=\columnwidth]{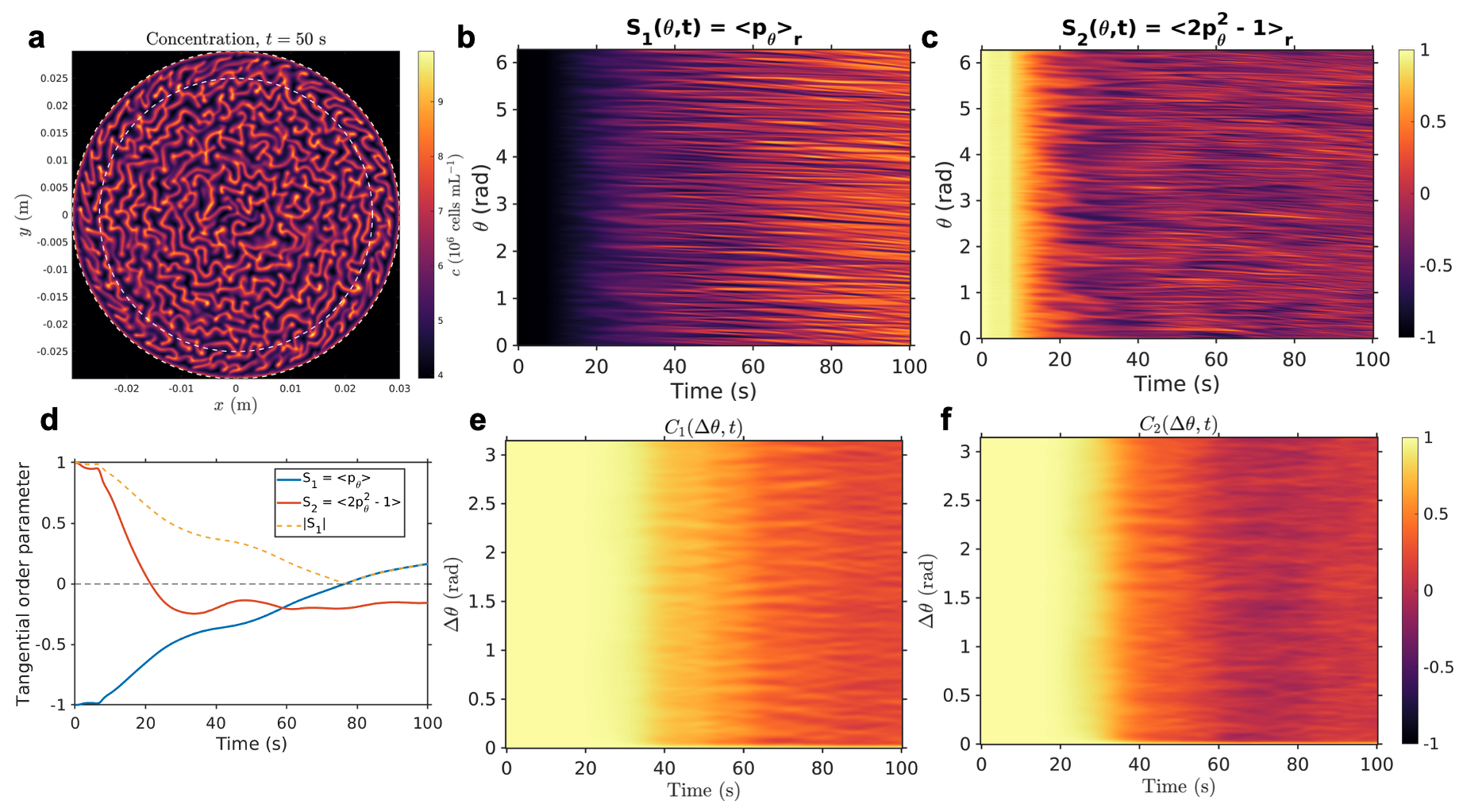}
	\caption{\textbf{Breakup of tangential polar order during the
			activity-driven instability.}
		\textbf{a,} Simulated cell-concentration field at
		\(t=50~\mathrm{s}\). The white dashed circles delimit the annular
		region used for the orientational analysis. The concentration is
		shown in units of \(10^6~\mathrm{cells\,mL^{-1}}\).
		\textbf{b,c,} Angularly resolved tangential polar and nematic order
		parameters,
		\(S_1(\theta,t)=\langle p_\theta\rangle_r\) and
		\(S_2(\theta,t)=\langle2p_\theta^2-1\rangle_r\), respectively,
		where
		\(p_\theta=\widehat{\mathbf P}\cdot\hat{\boldsymbol\theta}\)
		is the tangential component of the regularised unit polarisation and
		\(\langle\cdot\rangle_r\) denotes a radial average over the annulus
		in \textbf{a}. With positive \(p_\theta\) defined in the
		counter-clockwise direction, the initially clockwise tangential
		state has \(S_1\simeq-1\), whereas \(S_2\simeq1\) indicates strong
		tangential alignment irrespective of polarity. The emergence of
		angularly heterogeneous streaks shows the progressive fragmentation
		of the initially coherent polar texture.
		\textbf{d,} Annulus-averaged order parameters \(S_1(t)\) and
		\(S_2(t)\). The reduction in \(\lvert S_1\rvert\) and \(S_2\)
		indicates loss of coherent tangential polarity and alignment as
		radial and distorted polarisation components develop.
		\textbf{e,f,} Time-resolved angular polar and nematic correlations,
		\(C_1(\Delta\theta,t)\) and \(C_2(\Delta\theta,t)\), computed in the
		local polar basis. Their progressive decay with angular separation
		and time indicates a reduction in orientational coherence and the
		breakup of the annular polarisation into spatially heterogeneous
		domains.}
	\label{fig:correlation}
\end{figure}

The polarisation field revealed the orientational structure underlying the
instability. Although the imposed polarisation remained tangential at the
outer boundary, the field immediately inside the boundary progressively
rotated inward, while predominantly outward-oriented domains emerged farther
into the chamber. Narrow regions of reduced polarisation magnitude,
\(|\mathbf P|\), formed where oppositely oriented domains met. By contrast,
bands of enhanced \(|\mathbf P|\) spatially coincided with the
high-cell-concentration ridges shown at the corresponding times in
Fig.~\ref{fig:model}a, with the local polarisation tending to follow the ridge
contours (Fig.~\ref{fig:polar_order}a,b). The polarisation-phase maps further
showed the fragmentation of the initially coherent polar texture into
spatially heterogeneous orientational domains. Concurrently, heterogeneous
cellular consumption and transport increasingly modulated the oxygen field,
transforming its initially predominantly radial profile into a complex,
time-dependent oxygen landscape as the corona-like pattern developed
(Fig.~\ref{fig:polar_order}c).

To quantify this orientational reorganisation, we analysed the simulated
polarisation within the annular region shown in
Fig.~\ref{fig:correlation}a. The polarisation was expressed in the local polar
basis, with tangential component
$
p_\theta
=
\widehat{\mathbf P}\cdot\hat{\boldsymbol\theta},
$
where \(\widehat{\mathbf P}\) is the regularised unit polarisation defined in
the Methods. We evaluated the angularly resolved tangential polar and nematic
order parameters,
$
S_1(\theta,t)=\langle p_\theta\rangle_r$ and 
$
S_2(\theta,t)=\langle2p_\theta^2-1\rangle_r,
$
together with their annulus-averaged values. The averages were computed
without concentration weighting, so that they characterised the
orientational structure across the full selected annulus rather than
emphasising only its highest-concentration regions. With \(\hat{\boldsymbol\theta}\) defined in the counter-clockwise direction,
the initially clockwise tangential state had \(S_1\simeq-1\), whereas
\(S_2\simeq1\) indicated strong tangential alignment irrespective of polarity.
As the activity-driven instability developed, the magnitude of the
annulus-averaged \(S_1\) decreased and \(S_2\) was reduced, while both
quantities became strongly heterogeneous in angle
(Fig.~\ref{fig:correlation}b--d). The corresponding polar and nematic
correlations, \(C_1(\Delta\theta,t)\) and \(C_2(\Delta\theta,t)\), also
decayed over progressively shorter angular separations
(Fig.~\ref{fig:correlation}e,f), demonstrating the loss of long-range
orientational coherence. Thus, corona formation is accompanied by an active
reorganisation of the polarisation field along the annular interface,
consistent with a coupled splay--bend instability rather than a purely
density-driven deformation.

Taken together, the simulations separate three roles in the ring-to-corona
transition: oxygen-dependent taxis assembles and radially positions the
annular concentration band, dipolar active stresses destabilise the resulting
interface and generate protrusive deformation and flow, and the signed
tangential anchoring selects the direction of the long-time circulation.
\section*{Discussion}
We have shown that an oxygen field shaped by peripheral supply and cellular
consumption can assemble and destabilise an active interface in a living
polar fluid. In circular chambers with an air-exposed periphery, initially
homogeneous suspensions of \textit{E. gracilis} reorganised into an annular
cellular accumulation while a labyrinthine bioconvective pattern developed in
the chamber interior. The annulus was neither prescribed in the initial
condition nor imposed by a patterned boundary; it emerged from the coupled
effects of oxygen transport and consumption, oxygen-regulated motility and
bidirectional oxytaxis. Once established, the annular band developed
finite-wavelength protrusions and sustained collective circulation. The
experiments therefore support a two-stage mechanism: oxygen-dependent taxis
assembles and radially positions the cellular interface, whereas active
stresses subsequently destabilise it.

This interpretation extends earlier studies of oxygen-mediated responses in
\textit{Euglena}. Checcucci \textit{et al.} showed that red-light-induced
accumulation in green \textit{Euglena} is largely mediated by photosynthetic
oxygen production rather than by direct red-light phototaxis
\cite{checcucci1974red}. Colombetti and Diehn subsequently reported dynamic
ring patterns in thin \textit{Euglena} suspensions and attributed them to
chemosensory responses to dissolved oxygen, including responses of opposite
sign depending on the absolute oxygen concentration
\cite{colombetti1978chemosensory}. Experiments with an externally imposed
oxygen gradient further showed that dark-adapted \textit{E. gracilis} swim
towards higher oxygen concentrations and accumulate at a stationary oxygen
front when cellular consumption balances oxygen delivery
\cite{porterfield1997oxytaxis}. Our experiments recover this classical
oxygen-guided localisation but extend it into an active-fluid regime. Sealing
the chamber periphery suppressed formation of a comparable annular
accumulation, whereas a ring still formed near an oxygen source in a confined
\(100~\mu\mathrm{m}\)-high geometry but did not develop a rotating corona.
These controls indicate that oxygen access is required for annular
localisation, but that oxygen-guided accumulation alone is insufficient to
produce the protrusive, collectively flowing state.

The numerical model supports this separation between chemical localisation
and mechanical destabilisation. An oxygen-coupled polar active-fluid model,
adapted from the Giomi--Marchetti framework
\cite{giomi2012polar}, qualitatively reproduced the observed progression from
an initially homogeneous concentration field to an annular accumulation and
then to a protrusive, flowing pattern. Oxygen entered the model through two
biologically motivated couplings: a sign-changing oxytactic torque that
reorients the polarisation towards or away from oxygen gradients, and an
oxygen-dependent swimming speed that suppresses motility in oxygen-poor
regions. During the early localisation stage, these responses generated
convergent radial cell transport and assembled the annular band. As the
cellular and flow fields developed, heterogeneous oxygen consumption and
transport transformed the initially predominantly radial oxygen profile into
a complex, time-dependent field. Within the explored parameter range, the
corona-like instability appeared only when the dipolar activity exceeded a
threshold; below this onset, the oxytactically assembled annulus remained
approximately axisymmetric and exhibited neither pronounced protrusions nor
sustained coherent circulation. The model therefore identifies distinct
roles for oxygen and activity: the oxygen field determines where the cellular
interface forms, whereas active stresses determine whether that interface
deforms and flows.

The sign of the active stress has a direct physical interpretation. In the
convention used here, \(\alpha>0\) corresponds to contractile,
puller-type stresses, whereas \(\alpha<0\) corresponds to extensile,
pusher-type stresses \cite{giomi2012polar}. Pullers draw fluid inward along
their swimming axis and expel it laterally, whereas pushers drive the opposite
flow. Because \textit{E. gracilis} behaves, on average over a flagellar beat,
as an off-axis puller \cite{giuliani2021how}, the use of positive
\(\alpha\) is consistent with the known single-cell hydrodynamics. In the
simulations, increasing the contractile activity amplified distortions of the
annular polarisation field and generated the associated protrusive deformation
and flow.

Although the onset and strength of the circulation were activity driven, its
direction was not selected by the predominantly radial oxygen field. The
simulations imposed a signed clockwise tangential Dirichlet condition for the
polarisation at the outer boundary, thereby explicitly breaking the
clockwise--counter-clockwise symmetry. Reversing the initial bulk
polarisation while retaining the clockwise boundary polarity modified the
early transient but did not reverse the long-time circulation, which remained
clockwise. Conversely, reversing the boundary polarity together with a
compatible initial polarisation reversed the final flow direction. Thus, for
the boundary model and parameter regime considered here, the signed
tangential anchoring is the dominant selector of the asymptotic circulation
direction, whereas the initial bulk polarisation primarily affects the
transient dynamics. The instability and the resulting circulation emerge
dynamically, but their chirality is boundary selected rather than
spontaneously chosen. The absence of an oxygen-imposed chiral bias is
consistent with experiments in which different sectors of the same corona
rotate in opposite directions.

The simulated polarisation field provides an orientational view of the
instability. Although the polarisation remained tangential at the imposed
outer boundary, the field immediately inside the boundary progressively
rotated inward, while predominantly outward-oriented domains emerged farther
into the chamber. Regions where oppositely oriented domains met were
characterised by reduced polarisation magnitude, whereas bands of enhanced
\(\lvert\mathbf P\rvert\) spatially coincided with the high-concentration
cellular ridges. Within these ordered bands, the local polarisation tended to
follow the ridge contours. The phase and correlation analyses further showed
that the initially coherent tangential texture fragmented into angularly
heterogeneous domains as the corona-like pattern developed. In particular,
the reduction in tangential polar and nematic order, together with the decay
of angular orientational correlations, demonstrates that corona formation is
accompanied by a loss of long-range tangential coherence and the development
of radial, splay and bend distortions. These observations are consistent with
an active reorganisation of the polarisation along a curved cellular
interface rather than with a purely density-driven deformation.

The characteristic protrusion wavelength of approximately
\(5~\mathrm{mm}\) was obtained independently from real-space intensity
profiles and from the angular dynamic structure factor. This agreement
suggests that the wavelength is an intrinsic feature of the instability
rather than a transient feature of a particular measurement. Its value is
likely selected by a balance involving active forcing, polar elasticity,
viscous and confinement-mediated dissipation, oxygen-controlled accumulation
and the geometry of the annular interface. A linear stability theory of the
oxygen-selected active band would provide a natural route to predicting the
onset activity, selected wavelength and propagation rate, and to determining
how they depend on chamber geometry, oxygen consumption, swimming speed,
polar elasticity and anchoring. Because the present simulations use a signed
tangential Dirichlet condition, the quantitative threshold, circulation
direction and detailed polar structure should be interpreted as properties of
that boundary model rather than as boundary-independent predictions.

These results connect classical bioconvection with active-matter
hydrodynamics. In bioconvection, biased swimming, oxygen transport and density
stratification generate macroscopic flow
\cite{platt1961bioconvection,pedley1992hydrodynamic,hill2005bioconvection,
	hillesdon1996bioconvection,tuval2005bacterial}. In active polar fluids,
internally generated stresses destabilise orientationally ordered states and
produce spontaneous flow and complex travelling patterns
\cite{simha2002hydrodynamic,ramaswamy2010mechanics,
	marchetti2013hydrodynamics,hatwalne2004rheology,giomi2012polar}. The
\textit{E. gracilis} corona combines these mechanisms: oxygen transport,
consumption and taxis create a concentrated cellular interface, while active
stresses acting through the reorganised polarisation field destabilise that
interface and generate macroscopic flow. The system therefore provides an
experimentally accessible connection between oxygen-mediated taxis,
bioconvective transport and active interfacial hydrodynamics.

The corona also differs from many previously studied active interfaces.
Activity can generate interfacial waves, giant fluctuations, droplet
formation and protrusive morphologies at active or active--passive material
boundaries
\cite{soni2019stability,adkins2022dynamics,tayar2023controlling,
	xu2023geometrical,denk2020pattern}. In those systems, the interface is
typically supplied by phase separation, material contrast, confinement or an
externally imposed boundary. Here, by contrast, the interface is assembled by
the cells through their response to an oxygen field that they simultaneously
modify by consumption and transport. The chemical field therefore does more
than bias swimming: it localises the cellular accumulation and thereby
determines where active stresses are concentrated. This feedback suggests a
route towards chemically organised and potentially programmable active
interfaces in living fluids.

Several limitations define important directions for future work. The
spatiotemporal oxygen field is inferred from the chamber geometry, control
experiments and numerical model; direct oxygen measurements would test the
predicted transition from an initially radial profile to a heterogeneous,
time-dependent landscape and would quantify the oxygen range associated with
cell localisation. Similarly, the local cell concentration within the
annular band was not calibrated directly from image intensity. Simultaneous
measurements of cell concentration, velocity and polarisation would permit a
more quantitative determination of the activity coefficient and its
dependence on cell density, oxygen availability and metabolic state.
The experiments are also three-dimensional, with finite chamber height,
vertical bioconvective motion and an unsupported air--liquid meniscus, whereas
the present model is two-dimensional and represents confinement through
effective hydrodynamic parameters. Extending the theory to include vertical
structure, oxygen exchange through the meniscus and free-surface dynamics
will be necessary for a fully quantitative description. More generally,
alternative polarisation boundary conditions should be examined to determine
which features of the instability are robust and which depend on the imposed
anchoring.

The long-time evolution points to additional slow processes beyond the
initial ring-to-corona transition. Over periods exceeding two days, the
coherently rotating corona lost global rotational order and evolved into
larger protrusions exhibiting lateral oscillations, branching and
spatially heterogeneous motion. These secondary dynamics may reflect gradual
changes in oxygen availability, cell physiology, metabolism, crowding,
nutrient concentration or the accumulation of secreted products. Describing
this regime will require coupling active hydrodynamics to slower biological
and chemical variables that are not included in the present model.

More broadly, our results show that an environmental field can determine not
only how active matter moves, but also where its internally generated stresses
become concentrated. Peripheral oxygen supply and cellular consumption
assemble an annular living interface; active stresses destabilise that
interface and generate finite-wavelength protrusions and collective flow; and
the signed polar anchoring selects the direction of the resulting
circulation. This sequence provides a general mechanism by which metabolism,
taxis, orientational order and active hydrodynamics can cooperate to organise
living fluids, and suggests strategies for controlling active-matter
interfaces through chemical and geometric design.



\bibliography{bibliography}
\bibliographystyle{Science}
\section*{Acknowledgments}
We thank J. F. Joanny, J. Yeomans and I. S. Aronson for insightful discussions, and T. Malavath, M. Ali and Y. Liang for technical support. We also thank the Core Technology Platform at NYU Abu Dhabi, particularly V. Dhanvi and J. Govindan for fabrication of the PMMA moulds, Dr. Rezgui for microscopy support and Dr. Zhang for assistance with microfabrication. We acknowledge the NYU Abu Dhabi High-Performance Computing facility for
computational resources and data storage, which were critical to this work.
\section*{Author contributions}
A.G. conceived the study, designed the experiments, analysed the data, performed the numerical simulations with I.G. and wrote the initial manuscript. S.G. performed the experiments. A.G., S.V.R.A. and A.B. developed the analytical framework. All authors discussed the results and contributed to revision of the manuscript.
\section*{Data and materials availability}
All data needed to evaluate the conclusions of this study are included in the paper and/or the Supplementary Materials. Additional data are available from the corresponding author upon request.

\section*{Methods}
\renewcommand\thefigure{S\arabic{figure}}
\setcounter{figure}{0}
\subsubsection*{Cell culture}
\textit{Euglena gracilis} cells were maintained axenically in Hutner's
medium under standard laboratory growth conditions. Cultures were incubated
at \(25^{\circ}\mathrm{C}\) under a \(14\,\mathrm{h}:10\,\mathrm{h}\)
light--dark cycle with white-light irradiance of about
\(3\times10^{19}~\mathrm{photons\,m^{-2}\,s^{-1}}\), and culture
vessels were kept loosely capped to allow gas exchange. Cells were
subcultured into fresh sterile medium from actively growing cultures and
used for experiments during the motile growth phase. Before each experiment,
the suspension was gently mixed and diluted or concentrated in fresh medium
to the desired cell density. This light-grown, axenic protocol was chosen to
maintain highly motile, photosynthetically competent cells~\cite{wang2018euglena,
	yamashita2023method,atcc12894}.
\subsubsection*{Observation chamber}
Experiments were performed in a custom-made observation chamber fabricated from
polymethyl methacrylate (PMMA). The chamber consisted of two circular PMMA
plates forming the upper and lower boundaries of the sample. Each plate
contained a concentric groove fitted with a hydrophobic retaining ring at the
periphery. The plates were assembled coaxially, and their separation was
adjusted with screws to set the sample height to \(5~\mathrm{mm}\). Each retaining ring was
\(2~\mathrm{mm}\) high; therefore, at a \(5~\mathrm{mm}\) plate separation, the peripheral geometry left
an approximately 1 mm high unsupported air--liquid meniscus. This free-standing
meniscus formed the air-exposed boundary of the suspension and provided the
dominant route for oxygen exchange with the surrounding air. The chamber was filled with a suspension of \textit{E. gracilis} at a typical
density of \((5\text{--}8)\times10^6~\mathrm{cells\,mL^{-1}}\). During filling, the
suspension spread across the chamber and pinned at the hydrophobic retaining
rings rather than wetting beyond them. The suspension was introduced through a
small filling port at the centre of the upper plate, which was sealed after
filling to suppress oxygen exchange at the chamber centre. This ensured that
air contact occurred primarily through the peripheral meniscus.

\subsubsection*{Image acquisition and processing}
All experiments were conducted inside a dark enclosure to eliminate ambient
light gradients. During imaging and pattern development, the suspension was
illuminated uniformly from above using a red LED source centred at \(625~\mathrm{nm}\)
(LUMIMAX, LSR24-R). An optical diffuser was placed between the LED and the
sample to ensure spatially uniform illumination across the chamber. The photon
flux, measured at the position of the PMMA chamber, was maintained at
approximately \(5\times10^{17}~\mathrm{photons\,m^{-2}\,s^{-1}}\). This
illumination protocol was chosen to minimise phototactic steering:
conventional phototactic responses of \textit{Euglena} are dominated by
blue--green wavelengths, whereas red-light-induced accumulation in green cells
has been linked primarily to photosynthetic oxygen production rather than
direct phototaxis~\cite{checcucci1974red}. Oxygen-mediated ring patterns in
\textit{Euglena} have also been attributed to chemosensory responses to
dissolved oxygen~\cite{colombetti1978chemosensory}. Thus, spatially uniform red illumination minimised lateral phototactic cues
while allowing patterns to develop through oxygen transport, cellular motility,
bioconvection and active stresses.

For standard time-lapse experiments, more than \(10^4\) top-view images were
acquired at \(1~\mathrm{s}\) intervals with 8-bit intensity depth. For long-term
experiments, selected samples were monitored for up to \(5~\mathrm{d}\) to follow the
evolution of the corona. Low-frame-rate image sequences, when required, were
recorded at one frame every \(20~\mathrm{s}\). In the raw pseudo-dark-field images,
regions of high cell density appeared dark, whereas regions of low cell
density appeared bright.

Image preprocessing was performed in FIJI. Raw image stacks were inverted and
the temporal mean image was subtracted from each frame to enhance contrast and
reduce static background variations. In the processed images, high-density
regions appeared bright and were visualised using a pseudo-green colour map,
whereas low-density regions appeared dark. Space--time plots were constructed
by sampling the intensity along specified circular paths or annular regions
and stacking the resulting one-dimensional profiles in time. The characteristic
wavelength of the corona-like protrusions was obtained by Fourier analysis of
intensity profiles or kymographs using custom MATLAB scripts. The same
preprocessed image stacks were used for polar cross-correlation analysis of
the rotational dynamics.
\subsubsection*{Oxygen-coupled polar active-fluid simulations}

We modelled the \textit{E. gracilis} suspension as an oxygen-coupled polar
active fluid based on the hydrodynamic framework of Giomi and
Marchetti~\cite{giomi2012polar}. The equations were solved in the horizontal
plane, \(\mathbf r=(x,y)\), for the height-averaged cell concentration
\(c(\mathbf r,t)\), polarisation \(\mathbf P(\mathbf r,t)\), in-plane fluid
velocity \(\mathbf u(\mathbf r,t)\), pressure \(p(\mathbf r,t)\), and dissolved
oxygen concentration \(C(\mathbf r,t)\equiv C_{\mathrm O}(\mathbf r,t)\).
The concentrations \(c\) and \(C\) retain volumetric units of
\(\mathrm{mol\,m^{-3}}\). If
\[
n_{2\mathrm D}(\mathbf r,t)
=
\left\langle
\sum_n
\delta^{(2)}\!\left(\mathbf r-\mathbf r_n(t)\right)
\right\rangle
\]
is the coarse-grained areal number density, then
\begin{align}
	c(\mathbf r,t)
	&=
	\frac{n_{2\mathrm D}(\mathbf r,t)}{N_{\mathrm A}H},
	\label{eq:definition_c}
	\\
	\mathbf P(\mathbf r,t)
	&=
	\frac{1}{n_{2\mathrm D}(\mathbf r,t)}
	\left\langle
	\sum_n
	\hat{\mathbf n}_n(t)\,
	\delta^{(2)}\!\left(\mathbf r-\mathbf r_n(t)\right)
	\right\rangle ,
	\label{eq:definition_P}
\end{align}
where \(N_{\mathrm A}\) is Avogadro's constant, \(H\) is the chamber height,
\(\mathbf r_n(t)\) is the in-plane position of the \(n\)-th cell, and
\(\hat{\mathbf n}_n(t)\) is a unit vector along its instantaneous swimming
direction. The angular brackets denote coarse graining over a region large
compared with the cell size but small compared with the macroscopic pattern
scale.

The model incorporates two oxygen-dependent cellular responses. Oxygen
modulates the swimming speed and also reorients the polarisation. The
orientational response is positive oxytaxis at low oxygen, weak in an
intermediate range, and negative oxytaxis at high oxygen. By contrast, the
motility speed rises from a low value in oxygen-depleted regions to a plateau
at higher oxygen concentration.

The cell concentration obeys
\begin{equation}
	\partial_t c
	+
	\nabla\!\cdot\!
	\left[
	c\left(\mathbf u+w(C)\mathbf P\right)
	\right]
	=
	\nabla\!\cdot\!\left(\mathbf D\nabla c\right),
	\label{eq:cells_comsol}
\end{equation}
where
\begin{equation}
	D_{ij}
	=
	D_0\delta_{ij}
	+
	D_1P_iP_j
	\label{eq:diffusion_tensor_comsol}
\end{equation}
is the anisotropic diffusion tensor. Dissolved oxygen evolves through
advection, diffusion and cellular consumption,
\begin{equation}
	\partial_t C
	+
	\nabla\!\cdot(C\mathbf u)
	=
	D_{\mathrm O}\nabla^2C
	-
	k_{\mathrm O}cC,
	\label{eq:oxygen_comsol}
\end{equation}
where \(D_{\mathrm O}\) is the oxygen diffusivity and \(k_{\mathrm O}\) is
the bimolecular oxygen-consumption coefficient appearing in the locally
first-order sink \(k_{\mathrm O}cC\) for fixed cell concentration.

Following the reduced polar-fluid closure of Giomi and Marchetti, we set
\begin{equation}
	w_1
	=
	w_2
	=
	\frac{2\widetilde w_3}{\gamma}
	\equiv
	w(C),
	\label{eq:w_closure_comsol}
\end{equation}
so that the cell swimming speed, polar self-advection speed and local
splay--density coupling are governed by the same oxygen-dependent function.
The polarisation dynamics is
\begin{equation}
	\left[
	\partial_t
	+
	\left(\mathbf u+w(C)\mathbf P\right)\!\cdot\nabla
	\right]P_i
	+
	\Omega_{ij}P_j
	=
	\lambda E_{ij}P_j
	+
	\gamma^{-1}h_i
	+
	\zeta_{\mathrm O}
	S(C)
	\Pi^{\perp}_{ij}
	\partial_jC,
	\label{eq:polarization_comsol}
\end{equation}
with
\begin{equation}
	E_{ij}
	=
	\frac{1}{2}
	\left(
	\partial_i u_j+\partial_j u_i
	\right),
	\qquad
	\Omega_{ij}
	=
	\frac{1}{2}
	\left(
	\partial_i u_j-\partial_j u_i
	\right).
	\label{eq:strain_vorticity_comsol}
\end{equation}
The depth-integrated molecular field is
\begin{equation}
	h_i
	=
	-\left[
	a_2(c)+a_4(c)|\mathbf P|^2
	\right]P_i
	+
	K\nabla^2P_i
	-
	\frac{\gamma w(C)}{2}\,
	\frac{\partial_i c}{c_{\mathrm{safe}}},
	\label{eq:molecular_field_comsol}
\end{equation}
where
\begin{equation}
	c_{\mathrm{safe}}
	=
	\sqrt{c^2+c_{\mathrm{floor}}^2}
	\label{eq:csafe_comsol}
\end{equation}
regularises divisions by small concentration. The bulk coefficients are
\begin{equation}
	a_2(c)
	=
	a_0\frac{c^\ast-c}{c^\ast},
	\qquad
	a_4(c)
	=
	a_0\frac{c^\ast+c}{c^\ast}.
	\label{eq:bulk_coefficients_comsol}
\end{equation}
The two-dimensional coefficients are related to their three-dimensional
counterparts by
\[
K=K_{3\mathrm D}H,
\qquad
\gamma=\gamma_{3\mathrm D}H.
\]
Consequently, \(h_i\) has units of force per unit length.

The oxygen-induced reorientation is regularised through
\begin{equation}
	\Pi^{\perp}_{ij}
	=
	\delta_{ij}
	-
	\widehat P_i\widehat P_j,
	\qquad
	\widehat P_i
	=
	\frac{P_i}{\sqrt{|\mathbf P|^2+P_\ast^2}},
	\label{eq:projector_comsol}
\end{equation}
which approaches the transverse projector away from weakly polarised regions.
The sign-changing oxytactic response is
\begin{equation}
	S(C)
	=
	\frac{1}{2}
	\left[
	\tanh\!\left(
	\frac{C_- - C}{\Delta C}
	\right)
	-
	\tanh\!\left(
	\frac{C-C_+}{\Delta C}
	\right)
	\right].
	\label{eq:S_comsol}
\end{equation}
Thus,
\begin{align}
	S(C)&\simeq+1,
	&& C\ll C_-,
	\\
	S(C)&\simeq0,
	&& C_-\lesssim C\lesssim C_+,
	\\
	S(C)&\simeq-1,
	&& C\gg C_+.
\end{align}
For \(\zeta_{\mathrm O}>0\), this term turns the polarisation up oxygen
gradients at low oxygen and away from oxygen gradients at high oxygen.

The oxygen-dependent swimming speed is
\begin{equation}
	w(C)
	=
	w_{\min}
	+
	\frac{w_{\max}-w_{\min}}{2}
	\left[
	1+
	\tanh\!\left(
	\frac{C-C_-}{\Delta C}
	\right)
	\right].
	\label{eq:w_comsol}
\end{equation}
Thus,
\begin{align}
	w(C)&\simeq w_{\min},
	&& C\ll C_-,
	\\
	w(C)&\simeq
	\frac{w_{\min}+w_{\max}}{2},
	&& C\simeq C_-,
	\\
	w(C)&\simeq w_{\max},
	&& C\gg C_-.
\end{align}
Cells therefore slow in oxygen-depleted regions, accelerate around the lower
threshold \(C_-\), and remain motile at higher oxygen concentrations. Unlike
the orientational response \(S(C)\), the swimming speed does not decrease
again at high oxygen.

The incompressible momentum balance is
\begin{equation}
	\rho
	\left(
	\partial_t+\mathbf u\!\cdot\nabla
	\right)u_i
	=
	\partial_j\sigma_{ij}
	-
	\Gamma_{\mathrm d}u_i,
	\qquad
	\nabla\!\cdot\mathbf u=0,
	\label{eq:NS_comsol}
\end{equation}
where the Brinkman-type drag represents effective momentum loss to the upper
and lower confining plates,
\begin{equation}
	\mathbf f_{\mathrm d}
	=
	-\Gamma_{\mathrm d}\mathbf u,
	\qquad
	\Gamma_{\mathrm d}
	=
	s_{\mathrm d}
	\frac{\chi_{\mathrm d}\eta}{H^2}.
	\label{eq:confinement_drag_comsol}
\end{equation}
Here \(\eta\) is the dynamic viscosity, \(\chi_{\mathrm d}=12\) is the
parallel-plate resistance prefactor, and \(s_{\mathrm d}\) is a dimensionless
multiplier controlling the effective drag. For the drag-corrected simulations
reported here, \(H=5~\mathrm{mm}\) and \(s_{\mathrm d}=0.5\), giving
\[
\Gamma_{\mathrm d}
=
2.4\times10^2~\mathrm{Pa\,s\,m^{-2}}.
\]

The stress tensor is decomposed as
\begin{equation}
	\sigma_{ij}
	=
	2\eta E_{ij}
	+
	\sigma^r_{ij}
	+
	\sigma^a_{ij},
	\label{eq:stress_decomp_comsol}
\end{equation}
where
\begin{equation}
	\sigma^a_{ij}
	=
	\sigma^\alpha_{ij}
	+
	\sigma^\beta_{ij}.
	\label{eq:active_stress_decomp_comsol}
\end{equation}
Because the molecular field is depth integrated, the corresponding
three-dimensional reversible stress is
\begin{equation}
	\sigma^r_{ij}
	=
	-p\delta_{ij}
	+
	\frac{1}{H}
	\left[
	-\frac{\lambda}{2}
	\left(
	P_i h_j+P_j h_i
	\right)
	+
	\frac{1}{2}
	\left(
	P_i h_j-P_j h_i
	\right)
	-
	\bar{\lambda}
	\delta_{ij}
	\mathbf P\!\cdot\!\mathbf h
	\right].
	\label{eq:reversible_stress_comsol}
\end{equation}
The simulations used \(\bar{\lambda}=0\). The active stresses are
\begin{align}
	\sigma^\alpha_{ij}
	&=
	\alpha c
	\left(
	P_iP_j
	-
	\frac{1}{2}
	\delta_{ij}
	|\mathbf P|^2
	\right),
	\label{eq:alpha_stress_comsol}
	\\
	\sigma^\beta_{ij}
	&=
	\beta c
	\left(
	\partial_iP_j
	+
	\partial_jP_i
	+
	\delta_{ij}
	\nabla\!\cdot\mathbf P
	\right).
	\label{eq:beta_stress_comsol}
\end{align}
In the sign convention of Ref.~\cite{giomi2012polar}, \(\alpha>0\)
corresponds to contractile puller-type stress and \(\alpha<0\) to extensile
pusher-type stress. Because \textit{E. gracilis} behaves, on average over a
flagellar beat, as an off-axis puller~\cite{giuliani2021how}, we used
\(\alpha>0\). The polar-specific gradient active stress was disabled by
setting
\begin{equation}
	\beta=0.
	\label{eq:beta_zero_comsol}
\end{equation}
Polar symmetry nevertheless remains in the model through the self-advection
terms proportional to \(w(C)\), the sign-changing oxytactic torque and the
signed polarisation boundary condition. The active simulations shown in the
main text used
\[
\alpha
=
2\times10^8~\mathrm{N\,m\,mol^{-1}},
\]
whereas \(\alpha=0\) was used as an active-stress control.

The equations were solved in COMSOL Multiphysics 6.3 on a circular domain of
radius
$
R=30~\mathrm{mm}.
$
At the outer boundary, the total cell flux was set to zero,
\begin{equation}
	\widehat{\mathbf n}\!\cdot
	\left[
	c\left(
	\mathbf u+w\mathbf P
	\right)
	-
	\mathbf D\nabla c
	\right]
	=
	0,
	\label{eq:cell_noflux_comsol}
\end{equation}
and the dissolved oxygen concentration was fixed at its saturation value in
water,
\begin{equation}
	C\big|_{\partial\Omega}
	=
	C_{\mathrm O}^{\mathrm{sat,w}}
	\simeq
	0.28~\mathrm{mol\,m^{-3}}.
	\label{eq:oxygen_boundary_comsol}
\end{equation}
This oxygen condition represents rapid equilibration at the peripheral
air--liquid interface without explicitly resolving transport in the gas
phase. The same concentration scale was used to define \(C_-\), \(C_+\) and
\(\Delta C\).

The fluid satisfied an impermeable slip condition at the lateral boundary;
effective dissipation at the horizontal plates was represented by
\(\Gamma_{\mathrm d}\). A single pressure-point constraint fixed the
arbitrary pressure gauge and did not impose a physical pressure difference
across the chamber.

We imposed a signed tangential Dirichlet condition for the polarisation.
With \(\widehat{\mathbf n}=(n_x,n_y)\) denoting the outward unit normal,
we define
\begin{equation}
	\widehat{\mathbf t}_{s}
	=
	s_{\mathrm b}(-n_y,n_x),
	\qquad
	s_{\mathrm b}=+1
	\quad\text{(counter-clockwise)},
	\qquad
	s_{\mathrm b}=-1
	\quad\text{(clockwise)}.
	\label{eq:signed_tangent_comsol}
\end{equation}
The boundary condition is
\begin{equation}
	\left.\mathbf P\right|_{\partial\Omega}
	=
	P_{\mathrm{wall}}\widehat{\mathbf t}_{s}.
	\label{eq:tangential_anchoring_comsol}
\end{equation}
This constitutes hard anchoring in the mathematical sense because both
components of \(\mathbf P\) are prescribed, although the imposed magnitude
\(P_{\mathrm{wall}}=0.1\) is smaller than the locally preferred bulk
polarisation magnitude. In the production simulations, we set
\(s_{\mathrm b}=-1\), corresponding to the clockwise boundary condition
\begin{equation}
	P_x=P_{\mathrm{wall}}n_y,
	\qquad
	P_y=-P_{\mathrm{wall}}n_x.
\end{equation}
The opposite boundary polarity was obtained by changing the sign of
\(s_{\mathrm b}\).

The initial fields were
\begin{equation}
	c(\mathbf r,0)=c_0,
	\qquad
	C(\mathbf r,0)=C_{\mathrm O}^{\mathrm{sat,w}},
	\qquad
	\mathbf u(\mathbf r,0)=\mathbf 0.
	\label{eq:initial_scalar_flow_comsol}
\end{equation}
No annular concentration profile was imposed. The polarisation was initialised
in a regularised tangential texture. Defining
\begin{equation}
	r_{\varepsilon}
	=
	\sqrt{
		(x-x_c)^2+(y-y_c)^2+r_\ast^2
	},
	\qquad
	\mathbf t_{\varepsilon,s}
	=
	s_0
	\frac{
		\left(
		-(y-y_c),\,x-x_c
		\right)
	}{
		r_{\varepsilon}
	},
	\label{eq:regularized_tangent_initial}
\end{equation}
we used
\begin{equation}
	\mathbf P(\mathbf r,0)
	=
	P_{\mathrm{hom}}\,
	\mathbf R[\eta(\mathbf r)]\,
	\mathbf t_{\varepsilon,s},
	\label{eq:initial_P_comsol}
\end{equation}
where \(\mathbf R[\eta]\) denotes a planar rotation through the angle
\(\eta\), and
\begin{equation}
	P_{\mathrm{hom}}
	=
	\sqrt{
		\max\!\left(
		\frac{c_0-c^\ast}{c_0+c^\ast},
		0
		\right)
	}.
	\label{eq:initial_Pmag_comsol}
\end{equation}
The regularisation length
$
r_\ast=100~\mu\mathrm m
$
causes the polarisation magnitude to decrease smoothly towards zero at the
central \(+1\) defect. The angular perturbation was generated using a
zero-mean spatial random function with fixed seed, amplitude parameter
$
d\theta=0.346~\mathrm{rad},
$
and coordinate scale
$
L_c=150~\mu\mathrm m.
$
The sign \(s_0\) specifies the initial tangential polarity and was varied
independently of \(s_{\mathrm b}\) in the polarity-control simulations.

The cell and oxygen equations were implemented using the Transport of Diluted
Species interfaces, the polarisation equation using a General Form PDE
interface, and the momentum equation using Laminar Flow. For the cell field,
COMSOL's non-conservative transport form was supplied with the effective
velocity
\[
\mathbf u+w(C)\mathbf P
\]
and the compensating source
\[
-c\nabla\!\cdot
\left[
\mathbf u+w(C)\mathbf P
\right],
\]
which is algebraically equivalent to
Eq.~\eqref{eq:cells_comsol}. Consistent streamline and crosswind
stabilisation were enabled for the transport equations.

The domain was discretised using an unstructured triangular mesh. The global
mesh used a maximum element size of \(402~\mu\mathrm m\), a minimum element
size of \(1.2~\mu\mathrm m\), and a maximum growth rate of 1.05. Along the
outer boundary, a custom size restricted the maximum and minimum edge lengths
to \(100~\mu\mathrm m\) and \(10~\mu\mathrm m\), respectively, with a
maximum growth rate of 1.1. Boundary-layer elements were disabled for the
reported production runs.

Time integration used an implicit backward-differentiation formula with free
internal time stepping, BDF order 1--2, relative tolerance \(10^{-3}\), and a
maximum internal step of \(0.1~\mathrm s\). The coupled nonlinear system was
solved fully coupled using the direct MUMPS linear solver. Solutions were
stored at \(0.5~\mathrm s\) intervals. All dimensional parameter values are
listed in Table~S1.
\subsubsection*{Quasi-steady estimate of the oxygen-selected ring radius}
In the full numerical model, the radius and width of the annular
concentration band emerge dynamically from the coupled evolution of the cell
concentration, polarisation, fluid velocity and dissolved oxygen. We define
the instantaneous ring radius from the maximum of the azimuthally averaged
cell concentration,
\begin{equation}
	\overline{c}(r,t)
	=
	\frac{1}{2\pi}
	\int_0^{2\pi}
	c(r,\theta,t)\,\mathrm{d}\theta,
	\qquad
	r_{\mathrm{ring}}(t)
	=
	\underset{r}{\operatorname{arg\,max}}\,
	\overline{c}(r,t).
	\label{eq:ring_radius_definition}
\end{equation}
The following reduced calculation provides a leading-order interpretation of
the radial position selected during the early, approximately axisymmetric
stage of the dynamics.

The oxygen concentration generally increases from the chamber interior
towards the air-exposed outer boundary. In the model, the oxytactic
sensitivity satisfies
\begin{align}
	S(C)&>0,
	&& C\lesssim C_-,
	\\
	S(C)&\simeq0,
	&& C_-\lesssim C\lesssim C_+,
	\\
	S(C)&<0,
	&& C\gtrsim C_+.
\end{align}
For \(\zeta_{\mathrm O}>0\), cells in the low-oxygen region are therefore
reoriented towards increasing oxygen concentration, whereas cells in the
high-oxygen region are reoriented away from increasing oxygen concentration.
These two biases direct cells towards the intermediate oxygen band
\[
C_-\lesssim C\lesssim C_+.
\]
The ring is consequently expected to form near an isoconcentration contour
within this interval.

Because the oxytactic response is weak throughout the intermediate interval,
the model does not define a unique preferred oxygen concentration from
\(S(C)\) alone. We therefore introduce an effective concentration
\(C_{\mathrm{pref}}\in[C_-,C_+]\), defined as the oxygen concentration at the
maximum of the early annular cell profile,
\begin{equation}
	\overline{C}
	\bigl(r_{\mathrm{ring}}(t),t\bigr)
	\simeq
	C_{\mathrm{pref}},
	\label{eq:ring_radius_condition}
\end{equation}
where
\begin{equation}
	\overline{C}(r,t)
	=
	\frac{1}{2\pi}
	\int_0^{2\pi}
	C(r,\theta,t)\,\mathrm{d}\theta.
\end{equation}
Equivalently, \(C_{\mathrm{pref}}\) may be interpreted as the oxygen level at
which the azimuthally averaged radial cell flux changes sign. The precise
value generally depends on oxygen-dependent swimming, polar relaxation,
diffusion and the developing concentration profile. For a symmetric,
leading-order estimate one may use
\begin{equation}
	C_{\mathrm{pref}}
	\simeq
	\frac{C_-+C_+}{2},
	\label{eq:Cpref}
\end{equation}
but this midpoint is an approximation rather than an additional model
condition.

To obtain an analytical estimate, we consider a quasi-steady early-time
limit in which the cell concentration remains approximately homogeneous,
\(c(\mathbf r,t)\simeq c_0\), the flow is weak, and the oxygen field is
approximately radial. Neglecting oxygen advection and setting
\(\partial_tC\simeq0\), the oxygen equation reduces to
\begin{equation}
	D_{\mathrm O}
	\frac{1}{r}
	\frac{\mathrm{d}}{\mathrm{d}r}
	\left(
	r\frac{\mathrm{d}C}{\mathrm{d}r}
	\right)
	-
	k_{\mathrm O}c_0C
	=
	0,
	\label{eq:radial_oxygen_profile}
\end{equation}
where \(D_{\mathrm O}\) is the oxygen diffusivity and \(k_{\mathrm O}\) is
the bimolecular oxygen-consumption coefficient. For fixed \(c_0\), the
quantity \(k_{\mathrm O}c_0\) is the effective first-order oxygen-depletion
rate. Equation~\eqref{eq:radial_oxygen_profile} defines the oxygen
penetration length
\begin{equation}
	\ell_{\mathrm O}
	=
	\sqrt{
		\frac{D_{\mathrm O}}
		{k_{\mathrm O}c_0}
	}.
	\label{eq:oxygen_penetration_length}
\end{equation}

For a circular chamber of radius \(R\), with dissolved-oxygen concentration
\(C_b\) imposed at the air-exposed outer boundary and regularity at the
centre,
\begin{equation}
	C(R)=C_b,
	\qquad
	\left.
	\frac{\mathrm{d}C}{\mathrm{d}r}
	\right|_{r=0}
	=0,
	\label{eq:oxygen_boundary_conditions}
\end{equation}
the quasi-steady solution is
\begin{equation}
	C(r)
	=
	C_b
	\frac{
		I_0(r/\ell_{\mathrm O})
	}{
		I_0(R/\ell_{\mathrm O})
	},
	\label{eq:oxygen_bessel_solution}
\end{equation}
where \(I_0\) is the modified Bessel function of the first kind. In the
simulations,
\[
C_b
=
C_{\mathrm O}^{\mathrm{sat,w}},
\]
the saturation concentration of oxygen dissolved in water at the peripheral
air--liquid boundary.

Because \(I_0(x)\) increases monotonically for \(x\geq0\), the oxygen
concentration increases monotonically from
\begin{equation}
	C(0)
	=
	\frac{C_b}
	{I_0(R/\ell_{\mathrm O})}
\end{equation}
at the centre to \(C_b\) at the boundary. The quasi-steady estimate of the
ring radius is obtained from
\begin{equation}
	C(r_{\mathrm{ring}})
	=
	C_{\mathrm{pref}},
\end{equation}
giving
\begin{equation}
	r_{\mathrm{ring}}
	=
	\ell_{\mathrm O}
	\mathcal I_0^{-1}
	\left[
	\frac{C_{\mathrm{pref}}}{C_b}
	I_0\!\left(
	\frac{R}{\ell_{\mathrm O}}
	\right)
	\right].
	\label{eq:ring_radius_bessel}
\end{equation}
Here \(\mathcal I_0^{-1}\) denotes the functional inverse of \(I_0\) on
non-negative arguments, rather than its reciprocal.

The approximate inner and outer boundaries of the oxygen-selected band are
similarly defined by
\begin{equation}
	C(r_-)=C_-,
	\qquad
	C(r_+)=C_+,
	\label{eq:ring_edges_condition}
\end{equation}
which gives
\begin{align}
	r_-
	&=
	\ell_{\mathrm O}
	\mathcal I_0^{-1}
	\left[
	\frac{C_-}{C_b}
	I_0\!\left(
	\frac{R}{\ell_{\mathrm O}}
	\right)
	\right],
	\\
	r_+
	&=
	\ell_{\mathrm O}
	\mathcal I_0^{-1}
	\left[
	\frac{C_+}{C_b}
	I_0\!\left(
	\frac{R}{\ell_{\mathrm O}}
	\right)
	\right].
	\label{eq:ring_edges_bessel}
\end{align}
Since \(C(r)\) increases outwards, \(r_-<r_+\). A complete intermediate
oxygen band exists within the chamber only if
\begin{equation}
	\frac{C_b}
	{I_0(R/\ell_{\mathrm O})}
	\leq
	C_-
	<
	C_+
	\leq
	C_b.
	\label{eq:ring_band_existence_condition}
\end{equation}
More generally, an isoconcentration contour at \(C_{\mathrm{pref}}\) exists
when
\begin{equation}
	\frac{C_b}
	{I_0(R/\ell_{\mathrm O})}
	\leq
	C_{\mathrm{pref}}
	\leq
	C_b.
	\label{eq:ring_existence_condition}
\end{equation}

For a sufficiently narrow oxygen-selected interval, the radial width of the
band,
\[
\Delta r=r_+-r_-,
\]
can be estimated by linearising \(C(r)\) about a representative radius
\(r_0\), where \(C(r_0)=C_{\mathrm{pref}}\):
\begin{equation}
	\Delta r
	\simeq
	\frac{
		C_+-C_-
	}{
		\left|
		\partial_rC(r_0)
		\right|
	}.
	\label{eq:ring_width_general}
\end{equation}
From Eq.~\eqref{eq:oxygen_bessel_solution},
\begin{equation}
	\partial_rC(r)
	=
	\frac{C_b}{\ell_{\mathrm O}}
	\frac{
		I_1(r/\ell_{\mathrm O})
	}{
		I_0(R/\ell_{\mathrm O})
	},
	\label{eq:oxygen_gradient_bessel}
\end{equation}
where \(I_1\) is the modified Bessel function of the first kind. Hence
\begin{equation}
	\Delta r
	\simeq
	\ell_{\mathrm O}
	\frac{
		(C_+-C_-)
		I_0(R/\ell_{\mathrm O})
	}{
		C_b
		I_1(r_0/\ell_{\mathrm O})
	}.
	\label{eq:ring_width_bessel}
\end{equation}

In the strong-consumption limit,
\[
R\gg\ell_{\mathrm O},
\]
and within a distance of order \(\ell_{\mathrm O}\) from the outer boundary,
the large-argument expansion of the Bessel functions gives
\begin{equation}
	C(r)
	\simeq
	C_b
	\sqrt{\frac{R}{r}}\,
	\exp\!\left[
	-\frac{R-r}{\ell_{\mathrm O}}
	\right].
	\label{eq:oxygen_boundary_layer_curved}
\end{equation}
For \(R-r\ll R\), the curvature prefactor is close to unity and the leading
boundary-layer form becomes
\begin{equation}
	C(r)
	\simeq
	C_b
	\exp\!\left[
	-\frac{R-r}{\ell_{\mathrm O}}
	\right].
	\label{eq:oxygen_boundary_layer}
\end{equation}
The corresponding distance of the ring from the boundary is
\begin{equation}
	R-r_{\mathrm{ring}}
	\simeq
	\ell_{\mathrm O}
	\ln\!\left(
	\frac{C_b}{C_{\mathrm{pref}}}
	\right),
	\label{eq:ring_distance_boundary}
\end{equation}
or
\begin{equation}
	r_{\mathrm{ring}}
	\simeq
	R
	-
	\sqrt{
		\frac{D_{\mathrm O}}
		{k_{\mathrm O}c_0}
	}
	\ln\!\left(
	\frac{C_b}{C_{\mathrm{pref}}}
	\right).
	\label{eq:ring_radius_boundary_layer}
\end{equation}
Within this approximation, increasing \(c_0\) or \(k_{\mathrm O}\) decreases
the oxygen penetration length and shifts the selected band towards the
air-exposed boundary. Increasing \(D_{\mathrm O}\) increases the penetration
length and shifts the selected band farther into the chamber. At fixed
\(C_{\mathrm{pref}}\), increasing \(C_b\) also permits the corresponding
oxygen contour to extend farther into the chamber.

The reduced calculation contains no explicit dependence on the dipolar
activity coefficient \(\alpha\), supporting the interpretation that the
leading-order radial position of the early annulus is selected by oxygen
transport, consumption and taxis. This does not imply that the ring radius is
strictly independent of activity in the full nonlinear model. Once
collective flow, concentration heterogeneity and non-axisymmetric oxygen
transport develop, active stresses can shift, broaden and deform the annular
band indirectly through their effects on \(\mathbf u\), \(c\), \(\mathbf P\)
and \(C\).

Equations~\eqref{eq:ring_radius_bessel}--%
\eqref{eq:ring_radius_boundary_layer} should therefore be interpreted as
quasi-steady scaling estimates rather than exact predictions of the transient
ring radius. In the full simulations, \(r_{\mathrm{ring}}(t)\) is determined
from the evolving concentration field using
Eq.~\eqref{eq:ring_radius_definition}.
\subsection*{Angular order and correlations of the simulated polarisation field}

The simulated corona develops on an annular cellular interface. To quantify
the organisation of the polarisation along this interface, we expressed the
field in a local polar basis and calculated angular order parameters and
correlations within a selected annular region. This removes the purely
geometric rotation of the laboratory-frame basis: for example, a uniformly
tangential field has oppositely directed Cartesian vectors on opposite sides
of the chamber, although its orientation relative to the local tangent is
uniform.

At each stored time \(t_k\), the cell concentration and Cartesian
polarisation components,
\[
c(x,y,t_k),\qquad
P_x(x,y,t_k),\qquad
P_y(x,y,t_k),
\]
were exported from COMSOL on the same regular Cartesian grid. The
polarisation field is
\begin{equation}
	\mathbf P(x,y,t_k)
	=
	\left(
	P_x(x,y,t_k),
	P_y(x,y,t_k)
	\right).
\end{equation}
The analysis was restricted to the annulus
\begin{equation}
	r_{\mathrm{in}}
	\leq
	r
	\leq
	r_{\mathrm{out}},
	\qquad
	r
	=
	\sqrt{
		(x-x_c)^2+(y-y_c)^2
	},
	\label{eq:polar_annulus}
\end{equation}
where \((x_c,y_c)\) is the chamber centre. The Cartesian fields were
interpolated bilinearly onto a uniform polar sampling grid,
\begin{equation}
	x_{ij}
	=
	x_c+r_i\cos\theta_j,
	\qquad
	y_{ij}
	=
	y_c+r_i\sin\theta_j,
	\label{eq:polar_sampling_coordinates}
\end{equation}
with
\begin{equation}
	r_i
	\in
	[r_{\mathrm{in}},r_{\mathrm{out}}],
	\qquad
	\theta_j
	=
	\frac{2\pi j}{N_\theta},
	\qquad
	j=0,\ldots,N_\theta-1 .
\end{equation}
For the reported analysis, the polar grid contained \(N_r=120\) radial
samples and \(N_\theta=720\) angular samples. Samples for which the
interpolated fields were undefined were excluded.

The local radial and counter-clockwise tangential unit vectors are
\begin{equation}
	\widehat{\mathbf e}_r(\theta)
	=
	(\cos\theta,\sin\theta),
	\qquad
	\widehat{\mathbf e}_\theta(\theta)
	=
	(-\sin\theta,\cos\theta).
	\label{eq:local_polar_basis}
\end{equation}
We defined the regularised polarisation direction as
\begin{equation}
	\mathbf p(r,\theta,t)
	=
	\frac{
		\mathbf P(r,\theta,t)
	}{
		\sqrt{
			|\mathbf P(r,\theta,t)|^2+P_\ast^2
		}
	},
	\label{eq:regularised_orientation}
\end{equation}
where \(P_\ast\) is a small numerical regularisation. Away from regions of
vanishing polarisation, \(|\mathbf P|\gg P_\ast\), this approaches the unit
polarisation vector. Its radial and tangential components are
\begin{equation}
	p_r
	=
	\mathbf p\cdot\widehat{\mathbf e}_r,
	\qquad
	p_\theta
	=
	\mathbf p\cdot\widehat{\mathbf e}_\theta.
	\label{eq:polar_components}
\end{equation}
With this convention, \(p_\theta>0\) denotes counter-clockwise tangential
polarisation and \(p_\theta<0\) denotes clockwise tangential polarisation.

For any field \(A(r_i,\theta_j,t_k)\), the radial average at fixed
\((\theta_j,t_k)\) was calculated over the valid radial samples,
\begin{equation}
	\left\langle A\right\rangle_r
	(\theta_j,t_k)
	=
	\frac{1}{N_j(t_k)}
	\sum_{i\in\mathcal I_j(t_k)}
	A(r_i,\theta_j,t_k),
	\label{eq:radial_sample_average}
\end{equation}
where \(\mathcal I_j(t_k)\) is the set of valid radial indices and
\(N_j(t_k)=|\mathcal I_j(t_k)|\). These are uniform averages over the
discrete polar sampling grid; no concentration weighting or polar-coordinate
area factor was applied.

The angularly resolved tangential polar order parameter was defined as
\begin{equation}
	S_1(\theta,t)
	=
	\left\langle
	p_\theta(r,\theta,t)
	\right\rangle_r .
	\label{eq:S1_theta}
\end{equation}
This signed quantity distinguishes the two tangential polarities. A
well-ordered counter-clockwise tangential state gives \(S_1\simeq+1\),
whereas a clockwise tangential state gives \(S_1\simeq-1\).

The corresponding second-rank tangential order parameter was
\begin{equation}
	S_2(\theta,t)
	=
	\left\langle
	2p_\theta^2(r,\theta,t)-1
	\right\rangle_r .
	\label{eq:S2_theta}
\end{equation}
For a well-polarised field, \(S_2=1\) for tangential alignment of either
polarity, \(S_2=-1\) for radial alignment and \(S_2=0\) for an isotropic
distribution of orientations. Here \(S_2\) is used as a second-rank
diagnostic of the polar field; it does not imply that the underlying active
fluid has nematic symmetry.

The annulus-averaged order parameters were computed as averages over all
valid points of the polar sampling grid,
\begin{align}
	S_1(t)
	&=
	\frac{1}{N(t)}
	\sum_j
	\sum_{i\in\mathcal I_j(t)}
	p_\theta(r_i,\theta_j,t),
	\label{eq:S1_time}
	\\
	S_2(t)
	&=
	\frac{1}{N(t)}
	\sum_j
	\sum_{i\in\mathcal I_j(t)}
	\left[
	2p_\theta^2(r_i,\theta_j,t)-1
	\right],
	\label{eq:S2_time}
\end{align}
where
\begin{equation}
	N(t)
	=
	\sum_j N_j(t)
\end{equation}
is the total number of valid sampled points in the annulus.

To calculate angular correlations, we first formed the radially averaged
polarisation direction in the local polar basis,
\begin{equation}
	\overline{\mathbf p}_{\mathrm{ann}}(\theta_j,t)
	=
	\left(
	\langle p_r\rangle_r,
	\langle p_\theta\rangle_r
	\right).
	\label{eq:radially_averaged_orientation}
\end{equation}
Its magnitude,
\begin{equation}
	M(\theta_j,t)
	=
	\left|
	\overline{\mathbf p}_{\mathrm{ann}}(\theta_j,t)
	\right|,
\end{equation}
quantifies the coherence of the radial average. The corresponding normalised
mean orientation was
\begin{equation}
	\mathbf Q(\theta_j,t)
	=
	\frac{
		\overline{\mathbf p}_{\mathrm{ann}}(\theta_j,t)
	}{
		M(\theta_j,t)+\epsilon
	},
	\label{eq:mean_orientation_Q}
\end{equation}
where \(\epsilon\) prevents division by zero.

Angular positions at which the radial average was poorly defined were
downweighted using
\begin{equation}
	w_j(t)
	=
	N_j(t)\,
	M(\theta_j,t).
	\label{eq:angular_reliability_weight}
\end{equation}
For an angular separation
\(\Delta\theta_\ell=2\pi\ell/N_\theta\), the weight assigned to the pair
\((\theta_j,\theta_{j+\ell})\) was
\begin{equation}
	W_{j\ell}(t)
	=
	w_j(t)w_{j+\ell}(t),
	\label{eq:angular_pair_weight}
\end{equation}
with angular indices evaluated periodically modulo \(N_\theta\).

The time-resolved polar correlation was defined as
\begin{equation}
	C_1(\Delta\theta_\ell,t)
	=
	\frac{
		\displaystyle
		\sum_j
		W_{j\ell}(t)
		\,
		\mathbf Q(\theta_j,t)
		\cdot
		\mathbf Q(\theta_{j+\ell},t)
	}{
		\displaystyle
		\sum_j
		W_{j\ell}(t)
	}.
	\label{eq:C1_theta}
\end{equation}
The second-rank orientational correlation was
\begin{equation}
	C_2(\Delta\theta_\ell,t)
	=
	\frac{
		\displaystyle
		\sum_j
		W_{j\ell}(t)
		\left[
		2
		\left(
		\mathbf Q(\theta_j,t)
		\cdot
		\mathbf Q(\theta_{j+\ell},t)
		\right)^2
		-1
		\right]
	}{
		\displaystyle
		\sum_j
		W_{j\ell}(t)
	}.
	\label{eq:C2_theta}
\end{equation}
The first-rank correlation \(C_1\) distinguishes parallel and antiparallel
local polar orientations, whereas \(C_2\) treats them as equivalent.
Correlations were evaluated for
\[
0
\leq
\Delta\theta
\leq
\pi,
\]
because larger separations are redundant for a periodic angular field.

For a selected time interval \(t_a\leq t_k\leq t_b\), the time-averaged
correlations were calculated by accumulating the weighted pair numerators and
denominators over all selected frames,
\begin{equation}
	\overline C_1(\Delta\theta_\ell)
	=
	\frac{
		\displaystyle
		\sum_k
		\sum_j
		W_{j\ell}(t_k)
		\,
		\mathbf Q(\theta_j,t_k)
		\cdot
		\mathbf Q(\theta_{j+\ell},t_k)
	}{
		\displaystyle
		\sum_k
		\sum_j
		W_{j\ell}(t_k)
	},
	\label{eq:C1_time_average}
\end{equation}
and
\begin{equation}
	\overline C_2(\Delta\theta_\ell)
	=
	\frac{
		\displaystyle
		\sum_k
		\sum_j
		W_{j\ell}(t_k)
		\left[
		2
		\left(
		\mathbf Q(\theta_j,t_k)
		\cdot
		\mathbf Q(\theta_{j+\ell},t_k)
		\right)^2
		-1
		\right]
	}{
		\displaystyle
		\sum_k
		\sum_j
		W_{j\ell}(t_k)
	}.
	\label{eq:C2_time_average}
\end{equation}
Angular separations with negligible total pair weight were omitted from the
displayed correlation curves.

For comparison with physical distances along the annular interface, the
angular separation was converted to an arc length,
\begin{equation}
	s
	=
	r_{\mathrm{mid}}\Delta\theta,
	\qquad
	r_{\mathrm{mid}}
	=
	\frac{
		r_{\mathrm{in}}+r_{\mathrm{out}}
	}{2}.
	\label{eq:angular_arc_length}
\end{equation}
This conversion is a representative approximation for the finite-width
annulus.

With these definitions, \(S_1\) measures signed tangential polarity,
\(S_2\) measures tangential alignment independently of polarity,
\(C_1\) measures the angular persistence of local polar order and \(C_2\)
measures the persistence of second-rank orientational order. Because the
correlations are calculated after transforming each vector into its local
polar basis, they quantify the physical reorganisation of the polarisation
along the annular interface rather than the trivial rotation of the local
tangent around the circular geometry. Reversing the global convention for
the positive tangential direction reverses the sign of \(S_1\), but leaves
\(S_2\), \(C_1\) and \(C_2\) unchanged.
\subsubsection*{Polar-coordinate cross-correlation analysis of rotational motion}

The angular motion of the cellular corona was quantified directly from
time-lapse microscopy images using polar-coordinate cross-correlation. Each
image \(I(x,y,t_k)\) was converted to grayscale, represented in
floating-point format and rescaled to the intensity interval \([0,1]\).
The centre of the chamber, \((x_c,y_c)\), and the radial bounds of the
analysed annulus were determined from a representative image and held fixed
throughout each image sequence. The annular region was
\begin{equation}
	\mathcal A
	=
	\left\{
	(x,y):
	R_{\mathrm{in}}
	\leq
	\sqrt{(x-x_c)^2+(y-y_c)^2}
	\leq
	R_{\mathrm{out}}
	\right\}.
	\label{eq:rotation_annulus}
\end{equation}

For every frame, the image intensity within this region was interpolated onto
a regular polar grid,
\begin{equation}
	I_p(r_i,\theta_j,t_k)
	=
	I\!\left(
	x_c+r_i\cos\theta_j,\,
	y_c+r_i\sin\theta_j,\,
	t_k
	\right),
	\label{eq:polar_intensity_transform}
\end{equation}
where
\begin{equation}
	r_i\in[R_{\mathrm{in}},R_{\mathrm{out}}],
	\qquad
	\theta_j
	=
	\frac{2\pi j}{N_\theta},
	\qquad
	j=0,\ldots,N_\theta-1 .
\end{equation}
Intensities at non-integer image coordinates were obtained by bilinear
interpolation. The angular coordinate was defined to increase
counter-clockwise.

The frame-to-frame angular displacement was first estimated independently at
each sampled radius. To remove the angularly uniform component of the
intensity, we defined
\begin{align}
	A_i^k(\theta_j)
	&=
	I_p(r_i,\theta_j,t_k)
	-
	\left\langle
	I_p(r_i,\theta,t_k)
	\right\rangle_\theta,
	\\
	B_i^k(\theta_j)
	&=
	I_p(r_i,\theta_j,t_{k+1})
	-
	\left\langle
	I_p(r_i,\theta,t_{k+1})
	\right\rangle_\theta .
	\label{eq:mean_subtracted_profiles}
\end{align}
The circular cross-correlation between successive profiles was then
\begin{equation}
	C_i^k(m)
	=
	\sum_{j=0}^{N_\theta-1}
	A_i^k(\theta_j)
	B_i^k(\theta_{j+m}),
	\label{eq:angular_cross_correlation}
\end{equation}
where angular indices were evaluated modulo \(N_\theta\). The correlation was
computed using the fast Fourier transform,
\begin{equation}
	C_i^k
	=
	\mathcal F^{-1}
	\left[
	\mathcal F(A_i^k)^{*}
	\mathcal F(B_i^k)
	\right],
	\label{eq:angular_cross_correlation_fft}
\end{equation}
where \((\cdot)^{*}\) denotes complex conjugation.

Let
\begin{equation}
	m_i^\ast
	=
	\underset{0\leq m<N_\theta}{\operatorname{arg\,max}}\,
	C_i^k(m)
\end{equation}
denote the lag at the correlation maximum. The corresponding signed lag was
\begin{equation}
	\widetilde m_i^\ast
	=
	\begin{cases}
		m_i^\ast,
		&
		0\leq m_i^\ast\leq N_\theta/2,
		\\[3pt]
		m_i^\ast-N_\theta,
		&
		N_\theta/2<m_i^\ast<N_\theta,
	\end{cases}
\end{equation}
and the local angular displacement was
\begin{equation}
	\delta\theta_i^k
	=
	\widetilde m_i^\ast\Delta\theta,
	\qquad
	\Delta\theta
	=
	\frac{2\pi}{N_\theta}.
	\label{eq:local_angular_displacement}
\end{equation}
With the convention used here, positive
\(\delta\theta_i^k\) denotes counter-clockwise displacement and negative
\(\delta\theta_i^k\) denotes clockwise displacement.

To obtain one frame-to-frame displacement for the complete annulus, the
radial estimates were weighted by the non-negative correlation-peak height,
\begin{equation}
	w_i^k
	=
	\max
	\left[
	C_i^k(m_i^\ast),0
	\right].
	\label{eq:rotation_correlation_weight}
\end{equation}
The net displacement was calculated as the weighted circular mean,
\begin{equation}
	\Delta\Theta_k
	=
	\arg
	\left[
	\sum_i
	w_i^k
	\exp\!\left(
	\mathrm{i}\delta\theta_i^k
	\right)
	\right].
	\label{eq:net_angular_displacement}
\end{equation}
Frames for which the total valid weight vanished were treated as undefined
rather than being assigned zero displacement. Because the measured
frame-to-frame rotations were much smaller than \(\pi\), the circular mean
was equivalent to the corresponding weighted arithmetic mean while avoiding
ambiguities at the \(-\pi/\pi\) branch cut.

Defining
\begin{equation}
	\Delta t_k
	=
	t_{k+1}-t_k,
	\qquad
	t_{k+1/2}
	=
	\frac{t_k+t_{k+1}}{2},
\end{equation}
the frame-to-frame angular-velocity estimate was
\begin{equation}
	\Omega(t_{k+1/2})
	=
	\frac{\Delta\Theta_k}{\Delta t_k}.
	\label{eq:instantaneous_angular_velocity}
\end{equation}
Positive \(\Omega\) corresponds to counter-clockwise rotation and negative
\(\Omega\) to clockwise rotation. When the image times were expressed in
seconds, angular velocities were converted to
\(\mathrm{rad\,min^{-1}}\) by multiplying Eq.~\eqref{eq:instantaneous_angular_velocity}
by \(60\).

The cumulative angular displacement was defined at the original frame times
as
\begin{equation}
	\Theta_{\mathrm{cum}}(t_0)=0,
	\qquad
	\Theta_{\mathrm{cum}}(t_n)
	=
	\sum_{k=0}^{n-1}
	\Delta\Theta_k .
	\label{eq:cumulative_angular_displacement}
\end{equation}
Thus, a monotonically increasing cumulative displacement denotes sustained
counter-clockwise rotation, whereas a monotonically decreasing displacement
denotes sustained clockwise rotation.
%
%

%


\section*{Supplemental Information}
\renewcommand\thefigure{S\arabic{figure}}
\renewcommand\thetable{S\arabic{table}}
\setcounter{figure}{0}
\begin{figure}[!htbp]
	\centering
		\includegraphics[width=\columnwidth]{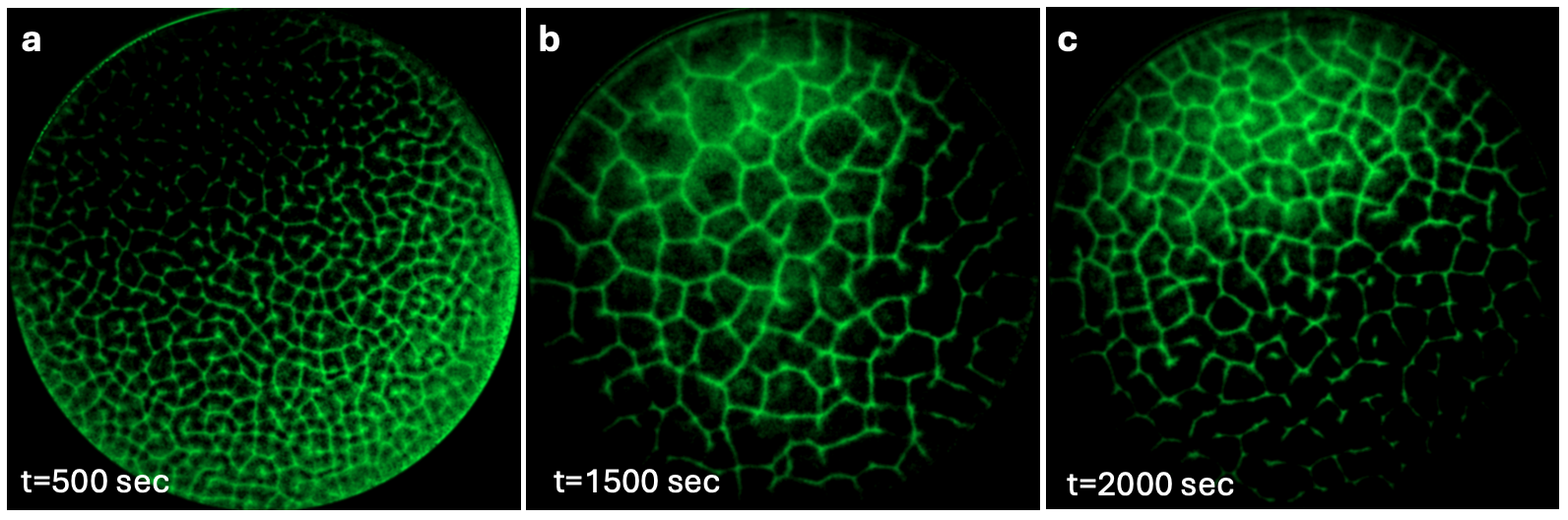}
		\caption{\textbf{Bioconvection patterns in a sealed air-impermeable chamber.}
			\textbf{(a--c)} Control experiment performed in a chamber with completely sealed boundaries, suppressing oxygen exchange with the surrounding air. Although dynamic bioconvection patterns are observed, no persistent ring-like cellular accumulation forms. This confirms that the annular ring observed in the main experiments requires an open peripheral air--liquid boundary and the associated radial oxygen gradient.}
		\label{fig:control}
\end{figure}
\begin{figure}[!htbp]
	\centering
	\includegraphics[width=\columnwidth]
	{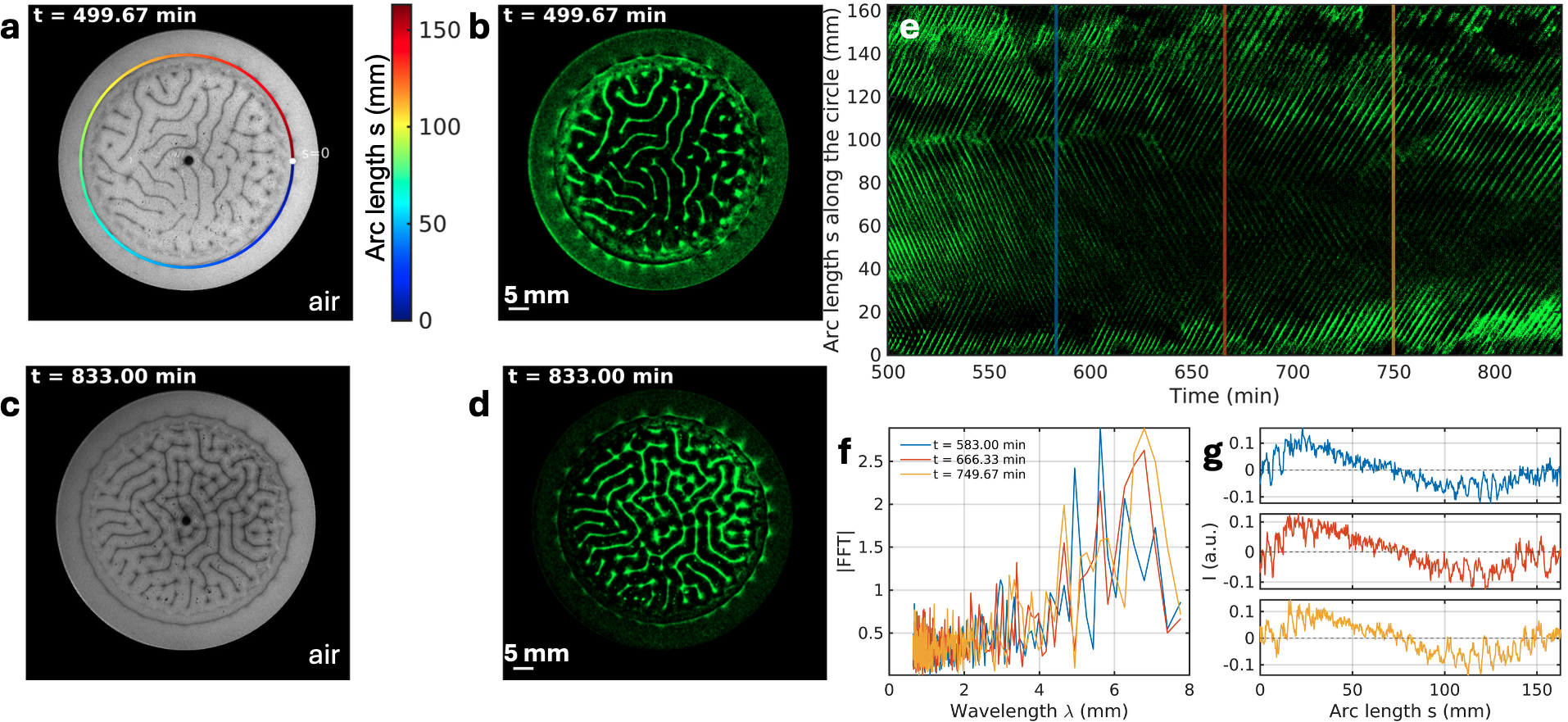}
	\includegraphics[width=0.96\columnwidth]
	{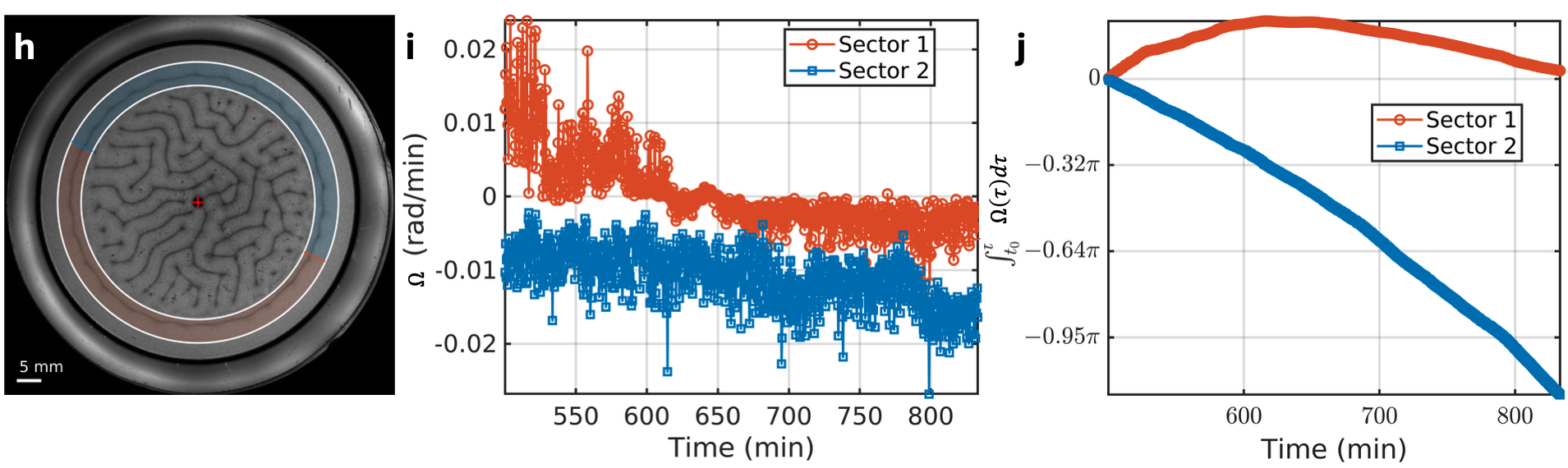}
	\caption{\textbf{Coexisting clockwise and counter-clockwise rotation in
			distinct sectors of the cellular corona.}
		\textbf{a,c,} Raw images of the corona at
		\(t=499.67\) and \(833.00~\mathrm{min}\), respectively. The coloured
		circular path in \textbf{a} defines the arc-length coordinate \(s\)
		used for the angular analysis; colour denotes distance along the
		path.
		\textbf{b,d,} Corresponding processed false-colour images
		highlighting the cellular ridges. Scale bars, \(5~\mathrm{mm}\).
		\textbf{e,} Space--time plot constructed by stacking intensity
		profiles sampled along the circular path in \textbf{a}. Streaks with
		opposite slopes in different arc-length sectors indicate simultaneous
		angular propagation in opposite directions. The vertical blue,
		orange and yellow lines mark the times analysed in \textbf{f,g}.
		\textbf{f,} Magnitude of the spatial Fourier transform of the
		intensity profiles in \textbf{g}, showing dominant finite-wavelength
		peaks of order \(5\text{--}7~\mathrm{mm}\).
		\textbf{g,} Representative intensity profiles at the three times
		indicated in \textbf{e}.
		\textbf{h,} Annular sectors used for polar-coordinate
		cross-correlation. The orange and blue arcs denote sectors 1 and 2,
		respectively; the white curves delimit the analysed annulus, and the
		red cross marks the centre of the polar transformation. Scale bar,
		\(5~\mathrm{mm}\).
		\textbf{i,} Instantaneous angular rotation rates
		\(\Omega(t)\) in the two sectors. With positive \(\Omega\) denoting
		counter-clockwise rotation, the opposite signs demonstrate
		counter-clockwise motion in sector 1 and clockwise motion in sector 2
		over a substantial portion of the observation interval. The
		counter-clockwise component weakens at later times, whereas the
		clockwise rotation remains persistent.
		\textbf{j,} Corresponding cumulative angular displacements,
		\(\Theta(t)=\int_{t_0}^{t}\Omega(\tau)\,\mathrm{d}\tau\). Sector 2
		accumulates a sustained clockwise displacement, whereas sector 1
		exhibits a smaller, non-monotonic counter-clockwise displacement,
		demonstrating spatially heterogeneous rotational dynamics within a
		single corona.}
	\label{fig:corona_BothDirections}
\end{figure}
\begin{table}[!htbp]
	\centering
	\scriptsize
	\setlength{\tabcolsep}{4pt}
	\renewcommand{\arraystretch}{1.4}
		\caption{\textbf{Parameters used in the COMSOL simulations.}
  Values are given in SI units. }
	\label{tab:comsol_parameters}
	\begin{tabular}{llll}
		\hline
		\textbf{Symbol} & \textbf{Description} & \textbf{Value} & \textbf{Unit} \\
		\hline
		\rowcolor{gray!15}
		\multicolumn{4}{l}{\textit{Geometry, initial state and anchoring}} \\
		\hline
		\(R\) & Circular-domain radius & \(3.0\times10^{-2}\) & \(\mathrm{m}\) \\
		\(c_0=2c^\ast\) & Initial homogeneous cell concentration & \(1.0\times10^{-11}\) & \(\mathrm{mol\,m^{-3}}\) \\
		\(c^\ast\) & Polar-order threshold concentration & \(5.0\times10^{-12}\) & \(\mathrm{mol\,m^{-3}}\) \\
		\(c_{\mathrm{floor}}=0.1c^\ast\) & Concentration regularisation floor & \(5.0\times10^{-13}\) & \(\mathrm{mol\,m^{-3}}\) \\
		\(d\theta\) & Initial angular-noise amplitude & \(0.346\) & \(\mathrm{rad}\) \\
		\(L_c\) & Angular-noise correlation length & \(1.5\times10^{-4}\) & \(\mathrm{m}\) \\
		\(r_{\mathrm{core}}\) & Initial polar-defect regularisation radius & \(1.0\times10^{-4}\) & \(\mathrm{m}\) \\
		\(P_{\mathrm{wall}}\) & Tangential boundary-polarisation magnitude & \(0.1\) & -- \\
		\(P_\ast\) & Polarisation regularisation parameter & \(1.0\times10^{-5}\) & -- \\
		\hline
		\rowcolor{gray!15}
		\multicolumn{4}{l}{\textit{Hydrodynamics and confinement}} \\
		\hline
		\(H\) & Chamber height & \(5.0\times10^{-3}\) & \(\mathrm{m}\) \\
		\(\rho_0\) & Base-fluid density in \(\rho=\rho_0+Mc\) & \(1.0\times10^{3}\) & \(\mathrm{kg\,m^{-3}}\) \\
		\(M\) & Effective cell molar mass in the density law & \(1.81\times10^{2}\) & \(\mathrm{kg\,mol^{-1}}\) \\
		\(\eta\) & Dynamic viscosity & \(1.0\times10^{-3}\) & \(\mathrm{Pa\,s}\) \\
		\(\chi_{\mathrm d}\) & Parallel-plate resistance prefactor & \(12\) & -- \\
		\(s_{\mathrm d}\) & Dimensionless confinement-drag multiplier & \(0.5\) & -- \\
		\(\Gamma_{\mathrm d}\) & Effective confinement-drag coefficient & \(2.4\times10^{2}\) & \(\mathrm{Pa\,s\,m^{-2}}\) \\
		\hline
		\rowcolor{gray!15}
		\multicolumn{4}{l}{\textit{Polar active-fluid parameters}} \\
		\hline
		\(D_0\) & Isotropic cell diffusivity & \(1.0\times10^{-8}\) & \(\mathrm{m^2\,s^{-1}}\) \\
		\(D_1\) & Anisotropic cell diffusivity & \(1.0\times10^{-8}\) & \(\mathrm{m^2\,s^{-1}}\) \\
		\(\lambda\) & Flow-alignment parameter (flow-tumbling regime) & \(0.1\) & -- \\
		\(\bar{\lambda}\) & Isotropic reversible coupling coefficient & \(0\) & -- \\
		\(K_{\mathrm{3D}}\) & Three-dimensional polar elastic coefficient & \(7.0\times10^{-12}\) & \(\mathrm{N}\) \\
		\(\gamma_{\mathrm{3D}}\) & Three-dimensional rotational friction & \(1.0\times10^{-3}\) & \(\mathrm{Pa\,s}\) \\
		\(K=K_{\mathrm{3D}}H\) & Height-integrated polar elastic coefficient & \(3.5\times10^{-14}\) & \(\mathrm{N\,m}\) \\
		\(\gamma=\gamma_{\mathrm{3D}}H\) & Height-integrated rotational friction & \(5.0\times10^{-6}\) & \(\mathrm{N\,s\,m^{-1}}\) \\
		\(\ell\) & Cellular polar length scale & \(1.0\times10^{-4}\) & \(\mathrm{m}\) \\
		\(a_0=K/\ell^2\) & Bulk polar-order scale & \(3.5\times10^{-6}\) & \(\mathrm{N\,m^{-1}}\) \\
		\(\tau_p=\ell^2\gamma/K\) & Polar relaxation time & \(1.43\) & \(\mathrm{s}\) \\
		\(\alpha\) & Dipolar active-stress coefficient & \(2.0\times10^{8}\) & \(\mathrm{N\,m\,mol^{-1}}\) \\
		\(\beta\) & Polar active-stress coefficient & \(0\) & \(\mathrm{N\,m^2\,mol^{-1}}\) \\
		\hline
		\rowcolor{gray!15}
		\multicolumn{4}{l}{\textit{Oxygen taxis and oxygen-dependent motility}} \\
		\hline
		\(C_{\mathrm O}^{\mathrm{sat,w}}\) & Saturation concentration of oxygen in water & \(2.8\times10^{-1}\) & \(\mathrm{mol\,m^{-3}}\) \\
		\(k_{\mathrm O}\) & Oxygen-consumption coefficient & \(1.0\times10^{10}\) & \(\mathrm{m^3\,mol^{-1}\,s^{-1}}\) \\
		\(\zeta_{\mathrm O}\) & Oxygen-reorientation coefficient & \(1.0\times10^{-2}\) & \(\mathrm{m^4\,mol^{-1}\,s^{-1}}\) \\
		\(w_{\max}\) & Maximum motility speed & \(1.0\times10^{-4}\) & \(\mathrm{m\,s^{-1}}\) \\
		\(w_{\min}=w_{\max}/5\) & Minimum motility speed & \(2.0\times10^{-5}\) & \(\mathrm{m\,s^{-1}}\) \\
		\(C_-=0.5C_{\mathrm O}^{\mathrm{sat,w}}\) & Lower threshold for positive oxytaxis & \(1.4\times10^{-1}\) & \(\mathrm{mol\,m^{-3}}\) \\
		\(C_+=0.8C_{\mathrm O}^{\mathrm{sat,w}}\) & Upper threshold for negative oxytaxis & \(2.24\times10^{-1}\) & \(\mathrm{mol\,m^{-3}}\) \\
		\(\Delta C=0.08C_{\mathrm O}^{\mathrm{sat,w}}\) & Smoothing width of the oxygen response & \(2.24\times10^{-2}\) & \(\mathrm{mol\,m^{-3}}\) \\
		\(D_{\mathrm O}\) & Oxygen diffusivity in water & \(1.98\times10^{-9}\) & \(\mathrm{m^2\,s^{-1}}\) \\
		\hline
	\end{tabular}
\end{table}

\end{document}